\documentclass{article}

\usepackage{times}
\usepackage{microtype}
\usepackage{graphicx}
\usepackage{subfigure}
\usepackage{booktabs}

\usepackage[skip=-5pt]{caption}
\makeatletter
\renewcommand{\paragraph}{%
  \@startsection{paragraph}{4}%
  {\z@}{0.5em}{-1em}%
  {\normalfont\normalsize\bfseries}%
}
\makeatother
\usepackage{hyperref}

\usepackage[accepted]{icml2024}

\usepackage{amsmath}
\usepackage{amsfonts}
\usepackage{amssymb}
\usepackage{mathtools}
\usepackage{amsthm}
\usepackage{tikz}
\usepackage[capitalize,noabbrev]{cleveref}

\theoremstyle{plain}

\theoremstyle{definition}

\theoremstyle{remark}

\usepackage{subcaption}
\usepackage{booktabs}
\usepackage{url}

\usepackage{interval}
\usepackage{tabularx}
\usepackage{color}
\usepackage{xspace}
\usepackage{mathtools}

\usepackage{multirow}
\usepackage{arydshln}
\usepackage[many]{tcolorbox} 
\usepackage{setspace}
\usepackage{lipsum,lmodern}
\AtBeginEnvironment{tcolorbox}{\small}

\usepackage[algo2e,ruled,vlined,linesnumbered]{algorithm2e}
\usepackage{wrapfig}

\usepackage[hang,flushmargin]{footmisc}
\newcommand\blfootnote[1]{
  \begingroup
  \renewcommand\thefootnote{}
  \footnote{#1}
  \addtocounter{footnote}{-1}
  \endgroup
}

\icmltitlerunning{Audio Flamingo}

\begin{document}

\twocolumn[
    \icmltitle{Audio Flamingo: A Novel Audio Language Model with Few-Shot Learning and Dialogue Abilities}

    \icmlsetsymbol{equal}{*}

    \begin{icmlauthorlist}
        \icmlauthor{Zhifeng Kong}{nv}
        \icmlauthor{Arushi Goel}{nv}
        \icmlauthor{Rohan Badlani}{nv}
        \icmlauthor{Wei Ping}{nv}
        \icmlauthor{Rafael Valle}{nv}
        \icmlauthor{Bryan Catanzaro}{nv}
    \end{icmlauthorlist}

    \icmlaffiliation{nv}{NVIDIA, Santa Clara, CA, USA}

    \icmlcorrespondingauthor{Zhifeng Kong}{zkong@nvidia.com}
    \icmlcorrespondingauthor{Wei Ping}{wping@nvidia.com}
    \icmlcorrespondingauthor{Rafael Valle}{rafaelvalle@nvidia.com}

    \icmlkeywords{Audio Language Model, Audio Understanding, In Context Learning, Retrieval Augmented Generation, Dialogue}
    \vskip 0.3in
]

\begin{abstract}
Augmenting large language models~(LLMs) to understand audio -- including non-speech sounds and non-verbal speech -- is critically important for diverse real-world applications of LLMs. 
In this paper, we propose \textbf{Audio Flamingo}, a novel audio language model with \textit{1)} strong audio understanding abilities, \textit{2)} the ability to quickly adapt to unseen tasks via in-context learning and retrieval, and \textit{3)} strong multi-turn dialogue abilities. We introduce a series of training techniques, architecture design, and data strategies to enhance our model with these abilities. Extensive evaluations across various audio understanding tasks confirm the efficacy of our method, setting new state-of-the-art benchmarks. Our demo website is \url{https://audioflamingo.github.io/} and the code is open-sourced at \url{https://github.com/NVIDIA/audio-flamingo}.
\blfootnote{~~$^1$NVIDIA, CA, USA. Correspondence to: Zhifeng Kong \texttt{<zkong@nvidia.com>}, Wei Ping \texttt{<wping@nvidia.com>}, Rafael Valle \texttt{<rafaelvalle@nvidia.com>}. \\
\\
\textit{Proceedings of the $\mathit{41}^{st}$ International Conference on Machine Learning}, Vienna, Austria. PMLR 235, 2024. Copyright 2024 by the author(s).
}
\end{abstract}

\section{Introduction} \label{sec:introduction}
The ability to understand sound is unarguably important and necessary for an agent to interact with the world. While large language models (LLMs) have shown an impressive ability to understand and reason about the world through text, their understanding of sound remains limited to transcriptions of speech \citep{lyu2023macaw}, thus making LLMs agnostic to important information in non-speech sounds and non-verbal speech. Even though recent contributions have improved their ability to understand sound \citep{gong2023ltu,lyu2023macaw,huang2023audiogpt,deshmukh2023pengi,chu2023qwen,tang2023salmonn}, there exists no model that combines: \textit{i)} strong audio understanding ability on various tasks \citep{deshmukh2023pengi}, \textit{ii)} the ability to execute multi-turn dialogues \citep{duan2023botchat}, and \textit{iii)} the ability to quickly adapt to unseen tasks without fine-tuning, for example, through in-context learning \citep{alayrac2022flamingo} and retrieval augmented generation \citep{lewis2020retrieval}. 

Our contribution to expand LLM's ability to understand sound is called \textbf{Audio Flamingo}: a novel audio language model that supports in-context learning (ICL), retrieval augmented generation (RAG), and multi-turn dialogues. It achieves state-of-the-art results on multiple audio understanding tasks.


Expanding LLM's ability to understand sound is a challenging task. 
The first challenge is extracting features from variable-length audio, and conditioning the LM on the audio features. While prior works have designed representations for audio of any length~\citep{wu2023large}, they can lose temporal information. In this work, we introduce an audio feature extractor with sliding window based on \citet{CLAP2023}, which we believe to capture temporal information better. To condition the LM on audio inputs, previous models prepended language tokens with audio tokens \citep{deshmukh2023pengi,chu2023qwen,tang2023salmonn}. This approach may have excessive overhead especially for long audio, as the complexity is quadratic with respect to the number of audio tokens. In contrast, we use cross attentions to fuse audio inputs into the LM in a similar way as Flamingo \citep{alayrac2022flamingo}. In our approach the complexity is linear in the number of audio tokens, thus making Audio Flamingo able to generalize to many audio inputs efficiently.

The second challenge is collecting and training on highly heterogeneous data. Prior works have collected and combined different datasets  for training \citep{doh2023lp,deshmukh2023pengi,chu2023qwen, gong2023ltu}. We follow their approach and curate a heterogeneous dataset with approximately 5.9 million audio-text pairs. Prior works have also designed different training curriculum, such as training on close-ended tasks first followed by open-ended tasks \citep{gong2023ltu,tang2023salmonn}. However, these result in a trade-off between close-ended and open-ended tasks, limiting the overall performance \citep{deshmukh2023pengi,tang2023salmonn}. We use a different approach based on a widely adopted and stable method to train LLMs \citep{ouyang2022training}. Specifically, we use two training stages: pre-training and supervised fine-tuning (SFT), each with different subsets and training techniques. These innovations make Audio Flamingo achieve the state-of-the-art results on several audio understanding benchmarks with $<\frac13$ number of parameters as \citet{chu2023qwen} and \citet{gong2023ltu}.

\begin{figure}[!t]
    \centering
    \includegraphics[trim={0 0 0 0}, clip, width=0.48\textwidth]{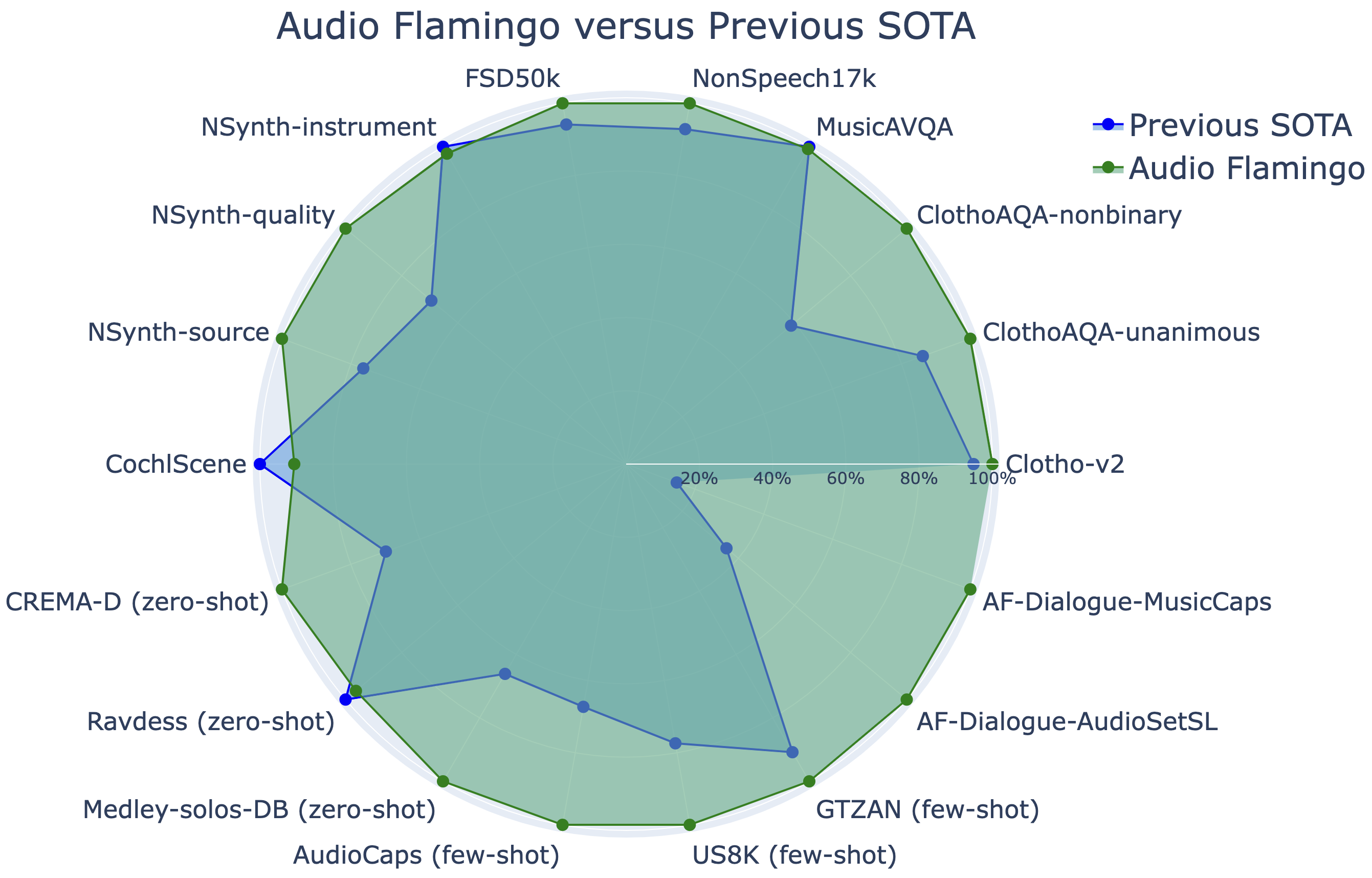}
    \vspace{-1em}
    \caption{Audio Flamingo versus previous state-of-the-art \citep{deshmukh2023pengi,chu2023qwen,gong2023joint,gong2023ltu,tang2023salmonn,ghosh2023recap} on a number of audio understanding benchmarks. The numbers are normalized such that the maximum of all models is $100\%$ on each task. Audio Flamingo sets the new state-of-the-art results on most of these tasks.}
    \label{fig: main radar}
\end{figure}

The third challenge is to give the audio language model the ability to quickly adapt to new tasks without fine-tuning, for instance, via in-context learning (ICL) \citep{brown2020language} and retrieval. While recent audio language models have shown zero-shot abilities \citep{deshmukh2023pengi,gong2023ltu}, they lack the ability to perform in-context few-shot learning to new tasks. In this paper, we introduce a series of techniques to realize this ability. We implement an efficient retrieval method, introduce an ICL template, and use retrieved samples to create interleaved ICL datasets for training. We also introduce a novel cross attention mask for interleaved samples. As a result, Audio Flamingo can be quickly adapted to new tasks via ICL and retrieval without task-specific fine-tuning. Our results confirm the efficacy of our approach and set new state-of-the-art few-shot benchmarks.

The last challenge is to give the audio language model the ability to chat with a user for many rounds. While prior methods have shown demos of dialogues \citep{gong2023ltu, chu2023qwen}, they lack systematic and quantitative evidence. To address this challenge, we create two multi-turn dialogue datasets with GPT-4 \citep{achiam2023gpt} based on detailed annotations of two datasets, with an emphasis on correlated context. 
We obtain a chat model by fine-tuning Audio Flamingo on these datasets. Our results show that our chat model has strong multi-turn dialogue ability and significantly outperforms previous methods.

We evaluate Audio Flamingo on a large and diverse set of close and open-ended benchmarks. A \textit{single} Audio Flamingo model surpasses the previous state-of-the-art on most benchmarks, and the \textit{chat} version of Audio Flamingo significantly outperforms baselines on dialogue benchmarks.  Figure~\ref{fig: main radar} summarizes the benchmark results of Audio Flamingo.
We also briefly discuss about the neural architecture and hyper parameters in the experiments. Our key contributions include:

\begin{enumerate}
\vspace{-1em}
\setlength\itemsep{0.0em}
\item We propose Audio Flamingo: a Flamingo-based audio language model for audio understanding with a series of innovations. Audio Flamingo achieves state-of-the-art results on several close-ended and open-ended audio understanding tasks.
\item We design a series of methodologies for efficient use of ICL and retrieval, which lead to the state-of-the-art few-shot learning results.
\item We enable Audio Flamingo to have strong multi-turn dialogue ability, and show significantly better results compared to baseline methods.
\end{enumerate}

\section{Related work} \label{sec:related_work}
\textbf{Multimodal LLMs.}
There has been tremendous progress in the area of multimodal LLMs. 
In addition to text, these models take inputs from various modalities such as vision \citep{tsimpoukelli2021multimodal,alayrac2022flamingo,yang2023re,driess2023palm,liu2023visual,li2023blip}, audio \citep{deshmukh2023pengi,gong2023joint,rubenstein2023audiopalm}, or multiple of them \citep{han2023imagebind,tang2023codi,moon2023anymal,zhao2023chatbridge}, and each has a different integration method. In the audio modality, prior works have looked at speech tasks \citep{chen2023salm,rubenstein2023audiopalm}, general audio understanding \citep{deshmukh2023pengi,gong2023ltu}, music understanding \citep{gardner2023llark,won2023foundation,li2023mert,liu2023music,doh2023lp}, or a combination of these \citep{gong2023joint,tang2023salmonn,chu2023qwen}. The focus of our paper is audio understanding, which includes non-speech sound and music, and non-verbal speech. Different from prior works, our model has stronger audio understanding ability, and is the first audio understanding model with \textit{i)} in-context few-shot learning ability, \textit{ii)} retrieval augmented generation ability, and \textit{iii)} strong multi-turn dialogue ability. 

\textbf{Audio encoders and representation.}
Many audio encoders extract audio features from the spectrogram, including CNN-based method \citep{kong2020panns} and Transformer-based methods \citep{gong2021ast,chen2022hts,defossez2022high,radford2023robust,gong2023whisper}. These methods are primarily targeted at solving a particular problem such as speech recognition or event detection. Based on these encoders, many joint audio-language embeddings have been proposed \citep{CLAP2022,CLAP2023,wu2023large,huang2022mulan,li2023mert}. These methods use contrastive learning to map audio and language embeddings into the same space, and are often trained on a large variety of audio and language. However, many of these methods compute a single embedding for an audio and therefore may lose temporal information. In this paper, we build an audio encoder with sliding windows based on ClapCap \citep{CLAP2023} to better capture long-range and temporal information.

\textbf{Data augmentation.}
Due to limited amount of high-quality human annotated sounds besides speech transcriptions, many works have proposed to augment textural description with existing LLMs such as GPT-4 \citep{achiam2023gpt}. A common strategy is to provide an LLM with annotated tags, timestamps, and other miscellaneous information, and then ask it to generate captions \citep{wu2023large, doh2023lp,mei2023wavcaps,gardner2023llark} or question-answering data pairs \citep{gong2023ltu,gong2023joint,liu2023music}. In this paper, we leverage existing LLMs to generate two multi-turn dialogue datasets based on detailed annotations, which enable our model strong dialogue abilities. 

\textbf{In-context learning (ICL).}
In-context learning is a kind of few-shot learning ability, where an LLM rapidly adapts to a desired task at inference time only after looking at a few examples in the prompt \citep{brown2020language}. It has widely shown success in natural language tasks \citep{wei2021finetuned} and visual-language tasks \citep{alayrac2022flamingo,yang2023re}. In the speech domain, ICL has been shown to help speech-related tasks such as speech recognition, translation, and processing \citep{gao2022wavprompt,wang2023can,hsu2023exploration,chen2023salm}. However, ICL for  general audio understanding is much less explored. In this paper, we propose the first audio understanding model with ICL ability.

\textbf{Retrieval-augmented generation (RAG).}
Retrieval-augmented generation for LLMs is to improve generation quality by using external knowledge, for example from an external database, which contains useful and related knowledge. It has been widely applied in natural language tasks \citep{guu2020retrieval,karpukhin2020dense,lewis2020retrieval,borgeaud2022improving} and visual-language models~\citep{yang2023re}. In the audio-language domain, \citet{ghosh2023recap} proposed a retrieval method for audio captioning by prepending captions from similar audios to the prompt. However, it does not provide the retrieved audio to the model. Consequently, the model loses information on how similar the retrieved audio is to the test audio. In contrast, we provide both the retrieved audio and text to our model. The benefit of this approach is that our model could determine when and how to use the retrieval based on the similarity between test and retrieved audio. We provide comparisons in our few-shot experiments.

\section{Method}\label{sec:methodology}
In this section, we introduce Audio Flamingo, an audio-understanding language model with few-shot learning via ICL and RAG. In Section \ref{sec: architecture}, we introduce the architecture used in Audio Flamingo, including the audio feature extractor, audio representation transformation layers, language model, and the conditioning method. In Section \ref{sec: training}, we introduce the training method of Audio Flamingo, including the training objective, design of masks, and training stages.

\subsection{Architecture}\label{sec: architecture}

\begin{figure*}[!t]
    \centering
    \includegraphics[width=0.8\textwidth]{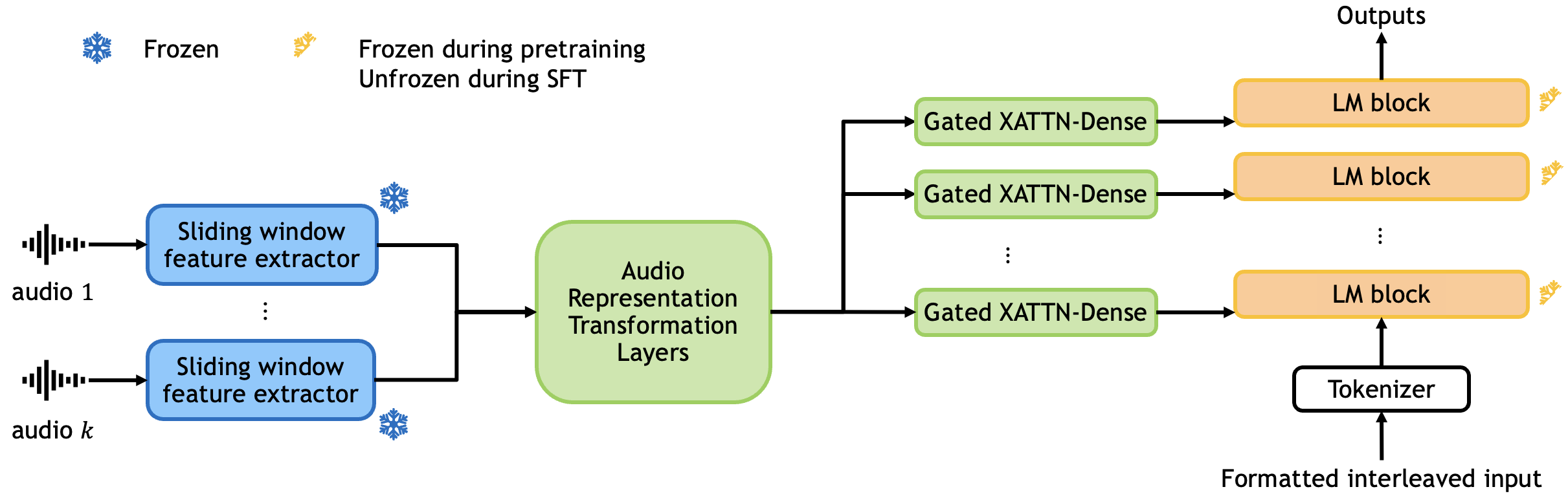}
    \vspace{-0.75em}
    \caption{Neural architecture of Audio Flamingo. It takes interleaved audio and text as input and outputs free-form text.}
    \vspace{-0.5em}
    \label{fig: architecture}
\end{figure*}

Our neural architecture is composed of four components: \textit{i)} an audio feature extractor with sliding window, \textit{ii)} audio representation transformation layers, \textit{iii)} a decoder-only language model, and \textit{iv)} gated xattn-dense layers. Figure~\ref{fig: architecture} summarizes the architecture.

\textbf{\textit{i)} Audio feature extractor with sliding window.}
We use ClapCap \citep{CLAP2023} as the audio feature extractor backbone, which we denote as $\mathcal{E}$. ClapCap is hard-coded to take 7-second of 44.1kHz raw audio as input, then transforms the audio into Mel-spectrogram of hop length 320, window length 1024, 64 Mel bins, and finally outputs a 1024-dimensional vector representation. 

We consider each 7-second segment as a window and use sliding windows to extract features for longer audio. The overlap between consecutive windows is $7\times0.75=5.25$ seconds. Formally, let $x(s\texttt{:}t)$ be the segment of $s$ to $t$ seconds in audio $x$. Then, the extracted feature is 
$\left[\mathcal{E}(x(0\texttt{:}7)), \mathcal{E}(x(\frac74\texttt{:}\frac{7\times5}{4})), \cdots, \mathcal{E}(x(\frac{7(m-1)}{4}\texttt{:}\frac{7(m+3)}{4})\right]$.
The intuition of this design is to capture long-range and temporal information that might be ignored in a single fused representation vector \citep{wu2023large}. We use a maximum of $m=16$ sliding windows, which supports a maximum of $33.25$ second audio length.~\footnote{We use $m=16$ because most training samples are $<30$s. Note that our cross attention complexity is linear in $m$ and therefore the audio length. In prior self-attention methods the attention complexity is quadratic in the audio length.}
Long audio will be cropped and short audio will be zero-padded. If an entire segment is zero-padded then we will mask the corresponding embedding at cross attention. If the input is interleaved data with $>1$ audio, we concatenate their sliding window representations.

\textbf{\textit{ii)} Audio representation transformation layers.}
We increase model capacity by further applying a few audio representation transformation layers to the concatenated audio feature representations described earlier. It is comprised of 3 self-attention layers \citep{vaswani2017attention}, with 8 heads and inner dimension 2048 each. This module is fully trainable.

\textbf{\textit{iii)} Language model.}
We use a decoder-only causal LM in our architecture. In this paper, we use \texttt{OPT-IML-MAX-1.3B} \citep{iyer2022opt}, a 1.3B parameter model with 24 LM blocks. It has been instruction-tuned on many natural language tasks. 

\textbf{\textit{iv)} Conditioning LM on audio representations.}
We use the {gated xattn-dense} layers from Flamingo \cite{alayrac2022flamingo} to achieve conditioning on audio inputs. Each layer has two blocks: 1) a residual block with cross attention and $\tanh$ gating, followed by 2) a residual block with dense layer and $\tanh$ gating. These layers are prepended to each LM block. 

\subsection{Training Method}\label{sec: training}

\textbf{\textit{i)} Training objective.}
Let $x$ be the mono-channel audio input, $y_{\mathrm{ins}}$ be the instruction text (e.g. question), and $y_{\mathrm{out}}$ be the output text. For conciseness we use $z=(x,y_{\mathrm{ins}},y_{\mathrm{out}})$ to represent each training sample.

We use maximum likelihood estimation (MLE) to train our model. Let $(y_{\mathrm{out}})_t$ be the $t$-th token and $(y_{\mathrm{out}})_{<t}$ the first $t-1$ tokens of the output. For a non-interleaved sample $z=(x,y_{\mathrm{ins}},y_{\mathrm{out}})$, the log-likelihood is 
\begin{equation}
\label{eq: loss single}
    \mathcal{L}(z)=\sum_{t=1}^{|y_{\mathrm{out}}|}\log p_{\theta}\left((y_{\mathrm{out}})_t|x,y_{\mathrm{ins}},(y_{\mathrm{out}})_{<t}\right).
\end{equation}
For an interleaved training sample composed of $J$ samples $z_{\mathrm{int}}=\{z^1,\cdots,z^J\}$, where $z^j=(x^j,y_{\mathrm{ins}}^j,y_{\mathrm{out}}^j)$, the log-likelihood is computed over all outputs:

\begin{equation}
\label{eq: loss interleaved}
\begin{array}{l}
    \displaystyle \mathcal{L}_{\mathrm{int}}(z_{\mathrm{int}}=\{z^1,\cdots,z^J\}) = \\
    \displaystyle ~\sum_{j=1}^J\sum_{t=1}^{|y_{\mathrm{out}}^j|}\log p_{\theta}\left((y_{\mathrm{out}}^j)_t|z^{<j},x^j,y_{\mathrm{ins}}^j,(y_{\mathrm{out}}^j)_{<t}\right).
\end{array}
\end{equation}

\begin{figure}[!t]
    \centering
    \includegraphics[width=0.5\textwidth]{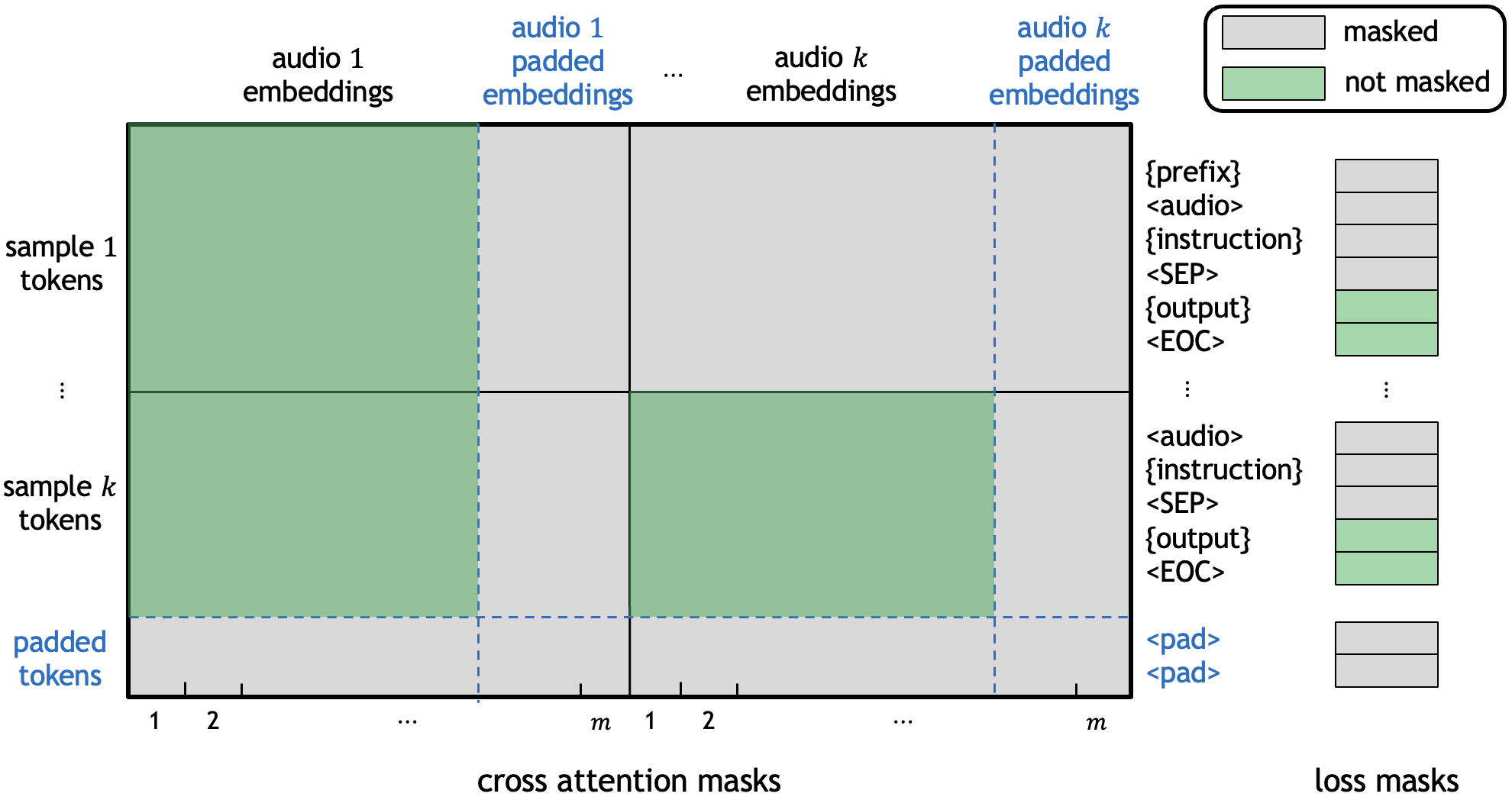}
    \vspace{-0.5em}
    \caption{Left: the block upper-triangular cross attention masks between text tokens and audio embeddings. Right: the loss mask of an interleaved training sample.}
    \label{fig: cross attention mask}
\end{figure}

Note this interleaved data loss is different from prior models, which compute losses only on the last output $y_{\mathrm{out}}^J$ \citep{yang2023re}, or have either none or indirect conditioning on prior multimodal inputs $x^{<j}$ \citep{alayrac2022flamingo,ghosh2023recap}. We expect \eqref{eq: loss interleaved} can help the model look at a various number (including zero when $j=1$) of ICL samples as well as the associated audio, thus improving robustness and training efficiency in a similar way as bucketing \citep{khomenko2016accelerating}, especially when the ICL samples are retrieved similar samples.
The corresponding loss mask is shown on the right-hand-side of Figure~\ref{fig: cross attention mask}.

Let $\{\mathcal{D}^i,i\in\mathcal{I}\}$ be all non-interleaved training datasets, and $\{\mathcal{D}_{\mathrm{int}}^{i'},{i'}\in\mathcal{I}_{\mathrm{int}}\}$ be all interleaved training datasets. The overall training objective is a weighted mixture of losses on each dataset:
\begin{equation}
    L=-\sum_{i\in\mathcal{I}} \lambda_i \mathbb{E}_{z\sim \mathcal{D}^i}\mathcal{L}(z) - \sum_{i'\in\mathcal{I_{\mathrm{int}}}} \lambda_{i'} \mathbb{E}_{z_{\mathrm{int}}\sim \mathcal{D}_{\mathrm{int}}^{i'}}\mathcal{L}_{\mathrm{int}}(z_{\mathrm{int}}),
\end{equation}
where $\lambda_i$'s are the weights for each dataset. The weights are constant hyper-parameters and have a huge impact on the final model. They are computed from the pre-defined number of epochs for each dataset (see Section \ref{sec: datasets} for the intuition, and Appendix \ref{appendix: dataset} for details).

\textbf{\textit{ii)} Cross attention masks.}
We use block upper-triangular cross attention masks for interleaved samples so that the likelihood of $j$-th output $p_{\theta}(y_{\mathrm{out}}^j)$ is conditioned only on the first $j$ audio inputs $x^{\leq j}$. We expect this helps making the model to look at previous audio. Figure~\ref{fig: cross attention mask} demonstrates the mask.

\textbf{\textit{iii)} Two training stages.}
We divide training into pre-training and supervised fine-tuning (SFT), a widely adopted and stable method in training LMs \citep{ouyang2022training}. During pre-training we only train the audio representation transformation layers and the {gated xattn-dense} layers. The purpose is to obtain a good set of initialization weights for these layers. During SFT, we unfreeze the entire LM, and train all modules but the audio encoder. \footnote{In initial experiments we found unfreezing the audio encoder caused training instability.} 

\section{Data}\label{sec:data}
\subsection{Datasets}
\label{sec: datasets}

In this section, we introduce our data strategies, including dataset collection, generation, and blending. We also introduce templates for each type of dataset.

\textbf{Dataset sources.}
We train our model on a variety of audio datasets that can be roughly classified into three types: music, non-speech general sound, and non-verbal speech. In this paper, we focus on these data given the immediate availability of state-of-the-art speech transcription models. We look at three types of tasks: (1) \textit{audio captioning} (CAP), where we would like the model to describe the audio in a sentence; (2) \textit{audio question-answering} (AQA), where we would like the model to answer questions regarding the audio, and (3) \textit{audio classification} (CLS), where we would like the model to classify the sound into one or more labels corresponding to events, scenes, music genres, instruments, qualities, and others. An overview of all training datasets is shown in Table \ref{tab: dataset by type}. 

\begin{table*}[!t]
    \small
    \centering
    \caption{All datasets used to train our model. The total number of audio-text pairs is approximately 5.9 million. The total length of audio is approximately 18.1 thousand hours.}
    \begin{tabular}{cccc}
        \toprule
        Audio Type & Task & Datasets & \#Audio-Text Pairs \\ \toprule
        \multirow{7}{*}{\shortstack{General\\Sound}} & \multirow{3}{*}{CAP} & WavCaps \citep{mei2023wavcaps}, Macs \citep{martin2021diversity}, & \multirow{3}{*}{$\sim$829 K} \\
        & & SoundDescs \citep{oncescu2021audio}, Clotho-v2 \citep{drossos2020clotho}, & \\
        & & WavText5K \citep{deshmukh2022audio}, LAION-630k \citep{wu2023large} & \\ \cline{2-4} 
        & AQA & Clotho-AQA \citep{lipping2022clotho}, Open-AQA \citep{gong2023joint} & $\sim$1970 K \\ \cline{2-4} 
        & \multirow{3}{*}{CLS} & AudioSet \citep{gemmeke2017audio}, FSD50k \citep{fonseca2021fsd50k}, & \multirow{3}{*}{$\sim$1091 K} \\
        & & CochlScene \citep{jeong2022cochlscene}, NonSpeech7K \citep{rashid2023nonspeech7k}, & \\
        & & Chime-Home \citep{foster2015chime}, Sonyc-UST \citep{cartwright2019sonyc} & \\ \hline
        
        \multirow{4}{*}{Music} & CAP & LP-MusicCaps \citep{doh2023lp}, MusicCaps \citep{agostinelli2023musiclm} & $\sim$1389 K \\ \cline{2-4} 
        & AQA & MusicQA \citep{liu2023music}, MusicAVQA \citep{li2022learning} & $\sim$94 K \\ \cline{2-4} 
        & \multirow{2}{*}{CLS} & NSynth \citep{nsynth2017}, MTG-Jamendo \citep{bogdanov2019mtg}, & \multirow{2}{*}{$\sim$459 K} \\
        & & FMA \citep{defferrard2016fma}, MusDB-HQ \citep{musdb18-hq}, & \\ \hline

        \multirow{3}{*}{Speech} & \multirow{3}{*}{CLS} & MSP-Podcast \citep{lotfian2017building}, Emov-DB \citep{adigwe2018emotional} & \multirow{3}{*}{$\sim$92 K} \\
        & & JL-Corpus \citep{james2018open}, Tess \citep{tess}, & \\
        & & MELD \citep{poria2018meld}, OMGEmotion \citep{barros2018omg} & \\ \hline
    \end{tabular}
    \label{tab: dataset by type}
\end{table*}

\begin{figure}[!t]
    \centering
    \includegraphics[width=0.5\textwidth]{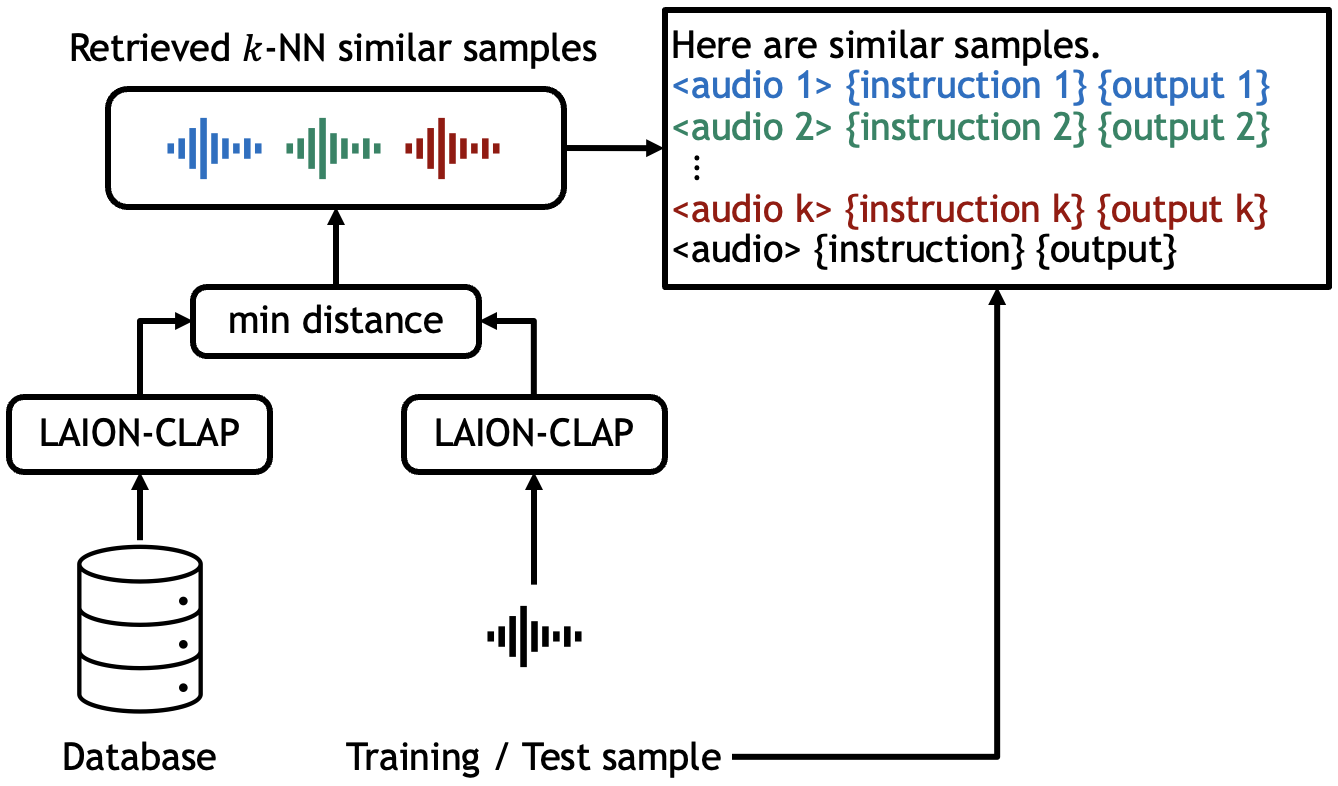}
    \vspace{-0.5em}
    \caption{Construction of ICL samples based on RAG. We use LAION-CLAP to find top-$k$ most similar samples from the database, and use the retrieved audio and text to construct an ICL training sample.}
    \label{fig: RAG}
\end{figure}

\textbf{ICL datasets.}
In order to give our model in-context learning and retrieval augmentation abilities, we construct ICL datasets for each of the raw datasets based on $k$NN computed on audio embeddings. Let $\mathcal{D}^i$ be the $i$-th training dataset. For each $z=(x,y_{\mathrm{ins}},y_{\mathrm{out}})\in\mathcal{D}^i$, we find its top-$k$ closest training samples in $\mathcal{D}^i$ excluding $z$, where the distance function is $\ell_2$ in the fused LAION-CLAP embedding space \citep{wu2023large} for the audio part of the sample. We use Faiss-gpu \citep{johnson2019billion} to accelerate searching. Figure~\ref{fig: RAG} demonstrates this process.

\textbf{Dataset staging and blending.}
We use different datasets during the pre-training stage and the supervised fine-tuning (SFT) stage. The selection is based on data quality, diversity, source, and size as described below. 1) Data quality: low quality datasets, including those with low-quality or noisy audio, low-quality text, and inaccurate text annotation, are used for pre-training. 2) Data diversity: datasets with less diversity or strong biases in label distributions are used for pre-training. 3) Data sources: datasets containing AI-generated contents are mostly used for pre-training, whereas some high-quality subsets may be used for SFT. 4) Data sizes: very large datasets may be used both for pre-training and SFT. 5) ICL datasets are used in the SFT stage. 

We assign different weights $\lambda_i$ when sampling from different datasets based on their sizes, quality, and diversity. The weights are computed from the number of epochs for each dataset. The details of staging and weights can be found in Appendix \ref{appendix: dataset}.

\subsection{Templates}
\label{sec: templates}

Our templates are based on OPT-IML's template \citep{iyer2022opt} and Flamingo's multimodal template \citep{alayrac2022flamingo}. For a non-interleaved sample, the template is describe below.

\newcommand{\slashn}{\textcolor{blue}{$\backslash$n}}
\begin{tcolorbox}[boxsep=0.5pt,left=2pt,right=2pt,top=1pt,bottom=1pt,colback=gray!5!white,colframe=gray!75!black]
  \texttt{<s>\{task description\}<audio>\{instruction\}}
  
  \texttt{Options:\slashn- option\textsubscript{1}\slashn $\cdots$- option\textsubscript{m}}
  
  \texttt{<SEP>\{output\}<EOC></s>}
\end{tcolorbox}

In this template, \texttt{<audio>} is the special token that informs the language model the location of audio in the context. The \texttt{\{task description\}} is natural language that tells the language model which task it is handling, for example ``\textit{The task is event classification}''. The \texttt{\{instruction\}} is the language instruction such as a \textit{question}. The \texttt{options} sentence is to tell the language model all options for classification so that it can classify an audio by outputting free-form text. The \texttt{\{output\}} is the ground truth output that will be trained. The \texttt{<SEP>} token is a separator that indicates the end of instruction, and \texttt{<EOC>} is the end-of-chunk token that indicates the end of a sample. 
Below is the template for interleaved (ICL) samples with $k+1$ tuples of (audio, instruction, output).

\begin{tcolorbox}[boxsep=0.5pt,left=2pt,right=2pt,top=1pt,bottom=1pt,colback=gray!5!white,colframe=gray!75!black]
  \texttt{<s>\{task description\}Here are similar samples.}
  
  \texttt{<audio>\{instruction\textsubscript{1}\}<SEP>\{output\textsubscript{1}\}<EOC>}
  
  {\texttt{\small{$\cdots$}}}
  
  \texttt{<audio>\{instruction\textsubscript{k}\}<SEP>\{output\textsubscript{k}\}<EOC>}
  
  \texttt{<audio>\{instruction\}}
  
  \texttt{Options:\slashn- option\textsubscript{1}\slashn $\cdots$- option\textsubscript{m}}
  
  \texttt{<SEP>\{output\}<EOC></s>}
\end{tcolorbox}

\subsection{Multi-Turn Dialogue Dataset}
\label{sec: dialogue dataset}

We aim at giving our model stronger abilities in dealing with complicated multi-turn dialogues. To achieve this, we use GPT-4 \citep{achiam2023gpt} to generate two multi-turn dialogue datasets. We construct these datasets based on the strongly labeled AudioSet-SL \citep{hershey2021benefit} and MusicCaps \citep{agostinelli2023musiclm}, with thresholding based on LAION-CLAP embeddings \citep{wu2023large} to filter undesirable samples. We name these two generated datasets \texttt{AF-Dialogue-AudioSetSL} and \texttt{AF-Dialogue-MusicCaps}, respectively. The detailed instructions, filtering method, dataset statistics, and examples are in Appendix \ref{appendix: dialogue}. We use the following template for an $s$-turn dialogue data sample.

\begin{tcolorbox}[boxsep=0.5pt,left=2pt,right=2pt,top=1pt,bottom=1pt,colback=gray!5!white,colframe=gray!75!black]
  \texttt{<s>The task is dialogue.<audio>}
  
  \texttt{user: \{instruction\textsubscript{1}\}}
  
  \texttt{assistant: <SEP>\{output\textsubscript{1}\}<EOC>}
  
  {\texttt{\small{$\cdots$}}}
  
  \texttt{user: \{instruction\textsubscript{s}\}}
  
  \texttt{assistant: <SEP>\{output\textsubscript{s}\}<EOC></s>}
\end{tcolorbox}

\section{Experiments}\label{sec:experiments}
In this section, we answer the following questions:\\
\textbf{Q1.} Does Audio Flamingo understand audio better than the state-of-the-art baselines?\\
\textbf{Q2.} How significantly does the ICL-based RAG help Audio Flamingo adapt to new tasks?\\
\textbf{Q3.} What is Audio Flamingo's ability to have multi-turn dialogues with a user?\\
\textbf{Q4.} Which specific configuration of Audio Flamingo works the best overall?

\subsection{Experimental Setup}
We use 8 NVIDIA A100 GPUs to train our model.
During pre-training, we use batch size $=384$, AdamW optimizer \citep{loshchilov2017decoupled} with learning rate $=1\times10^{-4}$ and weight decay $=0.1$. For efficiency, we use \textit{bf16} with automatic mixed precision. During supervised fine-tuning (SFT), we reduce the batch size to 128, the learning rate to $2\times10^{-5}$, and use \textit{fp32} for better numerical precision. We let the maximum number of interleaved samples to be 8 unless specified. We set the maximum number of text tokens to be 512. 

We compare to the most recent state-of-the-art baselines on several benchmarks. On each dataset, we choose the best score among all SOTA baselines as the reference value. Unless specified, we report accuracy for question-answering and single-label classification, F1 for multi-label classification, and CIDEr \citep{vedantam2015cider} for captioning and dialogues. Note we use free-form text output to evaluate our model at all times. We use a \textit{single} model to evaluate on all benchmarks except for dialogues, and a \textit{chat} model on dialogues.

For zero-shot and few-shot benchmarks, these datasets are excluded from the pre-training sets and SFT sets. For those derived from a parent dataset (e.g. AudioCaps audio are derived from AudioSet), we removed the training samples from the parent set as well as other child sets derived from that parent set.

\begin{table*}[!t]
    \centering
    \caption{Evaluation of Audio Flamingo versus SOTA baseline methods on in-distribution benchmarks. Reference values are the SOTA models for each task. Audio Flamingo shows strong audio understanding ability on captioning, question answering, and audio classification.}
    \begin{tabular}{lccll}
    \toprule
    Dataset & Task & Metric & Previous SOTA $\uparrow$ & Ours $\uparrow$ \\ \toprule 
    Clotho-v2 & CAP & CIDEr & $0.441$ \citep{chu2023qwen} & $\mathbf{0.465}$ \\
    ClothoAQA\textsubscript{unanimous} & AQA & ACC & $74.9\%$ \citep{chu2023qwen} & $\mathbf{86.9\%}$ \\
    ClothoAQA\textsubscript{non-binary} & AQA & ACC & $29.1\%$ \citep{deshmukh2023pengi} & $\mathbf{49.5\%}$ \\
    ClothoAQA\textsubscript{numerical} & AQA & ACC & $26.2\%$ \citep{deshmukh2023pengi} & $\mathbf{36.4\%}$ \\
    MusicAVQA\textsubscript{audio-only} & AQA & ACC & $\mathbf{72.1\%}$ \citep{chu2023qwen} & $71.6\%$ \\ 
    CochlScene & CLS & ACC & $\mathbf{91.6\%}$ \citep{deshmukh2023pengi} & $83.0\%$ \\
    NonSpeech7k & CLS & ACC & $79.0\%$ \citep{rashid2023nonspeech7k} & $\mathbf{85.1\%}$ \\
    FSD50k & CLS & F1\textsubscript{approx} & $65.6\%$ \citep{deshmukh2023pengi} & $\mathbf{69.7\%}$ \\
    NS\textsubscript{instrument} & CLS & ACC & $\mathbf{78.8\%}$ \citep{chu2023qwen} & $77.1\%$ \\
    NS\textsubscript{quality} & CLS & F1 & $46.3\%$ \citep{deshmukh2023pengi} & $\mathbf{66.7\%}$ \\
    NS\textsubscript{source} & CLS & ACC & $60.1\%$ \citep{deshmukh2023pengi} & $\mathbf{78.7\%}$ \\ 
    \bottomrule
    \end{tabular}
    \label{tab: in-domain results}
\end{table*}

\begin{table*}[!t]
    \centering
    \caption{Evaluation of Audio Flamingo versus SOTA baseline methods on zero-shot benchmarks. Reference values are the SOTA models for each task. Audio Flamingo shows strong zero-shot generalization ability.}
    \begin{tabular}{lcclc}
    \toprule
    Dataset & Task & Metric & Previous SOTA (0-shot) $\uparrow$ & Ours (0-shot) $\uparrow$ \\ \toprule 
    AudioCaps \citep{kim2019audiocaps} & CAP & CIDEr & $0.281$ \citep{salewski2023zero} & $\mathbf{0.502}$ \\
    CREMA-D \citep{cao2014crema} & CLS & ACC & $18.5\%$ \citep{deshmukh2023pengi} & $\mathbf{26.5\%}$ \\
    Ravdess \citep{livingstone2018ryerson} & CLS & ACC & $\mathbf{21.7\%}$ \citep{CLAP2023} & $20.9\%$ \\
    US8K \citep{salamon2014dataset} & CLS & ACC & $71.9\%$ \citep{deshmukh2023pengi} & $\mathbf{75.0\%}$ \\
    GTZAN \citep{sturm2013gtzan} & CLS & ACC & $\mathbf{71.0\%}$ \citep{han2023imagebind} & $67.9\%$  \\ 
    Medley-solos-DB \citep{lostanlen_2019_1344103} & CLS & ACC & $61.3\%$ \citep{deshmukh2023pengi} & $\mathbf{92.7\%}$ \\ 
    \bottomrule
    \end{tabular}
    \label{tab: zero shot results}
\end{table*}

\subsection{Q1: Strong Audio Understanding Ability}

We evaluate our model on several in-distribution (train-test) benchmarks, and compare to state-of-the-art audio language model baselines. The results are shown in Table \ref{tab: in-domain results}. Note that we define F1\textsubscript{approx} to measure inexact but similar predicted labels in FSD50k, where we consider the prediction to be correct if the sentence BERT similarity between output and ground truth is $>0.8$ \citep{reimers-2019-sentence-bert,reimers-2020-multilingual-sentence-bert}. This metric is applied to outputs from baselines as well.

Audio Flamingo can match or outperform SOTA baselines -- many of which are much larger LLMs (7B \citep{gong2023ltu,gong2023joint,chu2023qwen} or 13B \citep{tang2023salmonn}) -- on most tasks, indicating our proposed method has strong audio understanding ability. Our model also \textit{listens to the audio} better. On ClothoAQA, our model has $10.4\%$ higher accuracy than baselines on numerical question answering, indicating our model understands the number of occurrences better. On NSynth, our model has $20.4\%$ higher F1 on quality prediction and $18.6\%$ higher accuracy on source prediction, indicating our model understands the overall quality of audio better. In Appendix \ref{appendix: demo understanding}, we use qualitative samples to show that our model understands the order of appearance of different sounds, perceives loudness and its change over time, and perceives the distances of sounds from different objects. 

\subsection{Q2: In-Context Few-Shot Learning}

We aim to measure the effect of ICL-based RAG in Audio Flamingo when it is evaluated on unseen datasets. 

First, we report the results on several zero-shot benchmarks and comparison with SOTA zero-shot methods in Table \ref{tab: zero shot results}. The results indicate our method is better on most tasks and has strong generalization ability. 

We then apply ICL-based RAG to these benchmarks. We compare to our zero-shot results and the SOTA baseline of audio captioning on AudioCaps. The results on classification are shown in Table \ref{tab: few shot results}, and the comparison on retrieval-augmented audio captioning is shown in Table \ref{tab: audiocaps few shot results}. As expected, there is consistent improvement over zero-shot results, with an average improvement over $10\%$ for classification. Our method also significantly outperforms the SOTA retrieval-augmented audio captioning method on AudioCaps. In Appendix \ref{appendix: additional few-shot}, we show Audio Flamingo can adapt to unseen labels. In Appendix \ref{appendix: demo RAG}, we show Audio Flamingo \textit{looks at related retrieval} (e.g., by taking key words from retrieved captions), and \textit{ignores noisy retrieval}.

\subsection{Q3: Multi-Turn Dialogues}

We measure Audio Flamingo's ability to answer questions in a multi-turn dialogue setting. The context is more complex and strongly correlated between rounds (e.g. there exist many pronouns and follow-up questions). We fine-tune Audio Flamingo on the two sets that we generated (\texttt{AF-Dialogue-AudioSetSL} and \texttt{AF-Dialogue-MusicCaps}) to obtain a chat model. We evaluate the chat model on the test split of these two dialogue datasets. We take user instructions and let the model generate answers turn-by-turn (where previous generated answers become the chatting history for next generation). We compare to Qwen-Audio \citep{chu2023qwen}, LTU \citep{gong2023ltu}, and MU-LLaMA \citep{liu2023music} in Table~\ref{tab: multi-turn results}. \footnote{While the baseline methods claimed to support multi-turn dialogues, we were unable to find quantitative evidence.} Our chat model achieves significantly better results than baseline methods. In Appendix \ref{appendix: demo dialogue}, we use qualitative samples to show that our chat model \textit{captures context} such as prior information and pronouns better.

\begin{table}[!t]
    \centering
    \caption{Evaluation of few-shot results of Audio Flamingo with ICL-based RAG. $\Delta$ is the \textit{absolute} improvement of few-shot over zero-shot results in Table \ref{tab: zero shot results}. ICL-based RAG leads to consistent improvement over zero-shot results.}
    \begin{tabular}{lcc}
    \toprule
    Dataset & Ours (8-shot) $\uparrow$ & $\Delta $ $\uparrow$ \\ \hline 
    CREMA-D & $31.8\%$ & $5.3\%$ \\
    Ravdess & $35.2\%$ & $14.3\%$ \\
    US8K & $94.7\%$ & $19.4\%$ \\
    GTZAN & $79.5\%$ & $11.6\%$ \\ 
    Medley-solos-DB & $95.7\%$ & $3.0\%$ \\ \midrule
    Dataset & Ours (16-shot) $\uparrow$ & $\Delta $ $\uparrow$ \\ \hline 
    CREMA-D & $35.4\%$ & $8.9\%$ \\
    Ravdess & $40.2\%$ & $19.3\%$ \\
    US8K & $91.9\%$ & $16.9\%$ \\
    GTZAN & $76.3\%$ & $8.4\%$ \\ 
    Medley-solos-DB & $96.0\%$ & $3.3\%$ \\
    \bottomrule
    \end{tabular}
    \label{tab: few shot results}
\end{table}

\begin{table}[!t]
    \centering
    \vspace{-0.5em}
    \caption{Evaluation of retrieval-augmented audio captioning on AudioCaps. We compare Audio Flamingo to the SOTA baseline RECAP \citep{ghosh2023recap}. Audio Flamingo achieves significantly better results than RECAP.}
    \begin{tabular}{ccccc}
    \toprule
    Method & RECAP & Ours & Ours & Ours \\ \hline
    \# Shots & $4$ & $4$ & $8$ & $16$ \\ \hline
    CIDEr $\uparrow$ & $0.359$ & $0.518$ & $0.538$ & $\mathbf{0.546}$ \\ \bottomrule
    \end{tabular}
    \label{tab: audiocaps few shot results}
\end{table}

\begin{table}[!t]
    \centering
    \vspace{-0.5em}
    \caption{Evaluation of Audio Flamingo versus baseline methods on the multi-turn dialogue test sets. \texttt{A} stands for \texttt{AF-Dialogue-AudioSetSL}, \texttt{M} stands for \texttt{AF-Dialogue-MusicCaps}, and the superscript $\mathrm{H}$ stands for an additional held-out testset generated with \texttt{gpt-3.5-turbo}. We report CIDEr, Bleu4 \citep{papineni2002bleu}, and Rouge-L (R-L) \citep{lin2004rouge}. Methods with the $\dag$ superscript are fine-tuned on our dialogue training sets, and methods without $\dag$ are evaluated zero-shot on the dialogue test sets. Audio Flamingo significantly outperforms larger baseline models in all settings, indicating strong dialogue ability of our proposed model.}
    \begin{tabular}{ccccc}
    \toprule
    Testset & Method & CIDEr $\uparrow$ & Bleu4 $\uparrow$ & R-L $\uparrow$ \\ \midrule 
    \texttt{A} & Qwen-Audio & $0.507$ & $0.060$ & $0.292$ \\
    \texttt{A} & LTU & $0.580$ & $0.122$ & $0.324$ \\
    \texttt{A} & LTU$^{\dag}$ & $0.823$ & $0.153$ & $0.403$ \\
    \texttt{A} & Ours$^{\dag}$ & $\mathbf{1.622}$ & $\mathbf{0.237}$ & $\mathbf{0.473}$ \\ \hline 
    \texttt{A$^{\mathrm{H}}$} & LTU$^{\dag}$ & $0.523$ & $0.095$ & $0.343$ \\
    \texttt{A$^{\mathrm{H}}$} & Ours$^{\dag}$ & $\mathbf{1.904}$ & $\mathbf{0.219}$ & $\mathbf{0.476}$ \\ \midrule 
    \texttt{M} & MU-LLaMA & $0.585$ & $0.083$ & $0.348$ \\
    \texttt{M} & LTU & $0.168$ & $0.065$ & $0.217$ \\
    \texttt{M} & LTU$^{\dag}$ & $0.419$ & $0.108$ & $0.336$ \\
    \texttt{M} & Ours$^{\dag}$ & $\mathbf{1.143}$ & $\mathbf{0.142}$ & $\mathbf{0.417}$ \\ \hline 
    \texttt{M$^{\mathrm{H}}$} & LTU$^{\dag}$ & $0.558$ & $0.083$ & $0.347$ \\
    \texttt{M$^{\mathrm{H}}$} & Ours$^{\dag}$ & $\mathbf{1.350}$ & $\mathbf{0.207}$ & $\mathbf{0.448}$ \\
    \bottomrule
    \end{tabular}
    \label{tab: multi-turn results}
\end{table}

\begin{figure}[!t]
\centering
\begin{subfigure}
    \centering
    \includegraphics[width=0.48\textwidth]{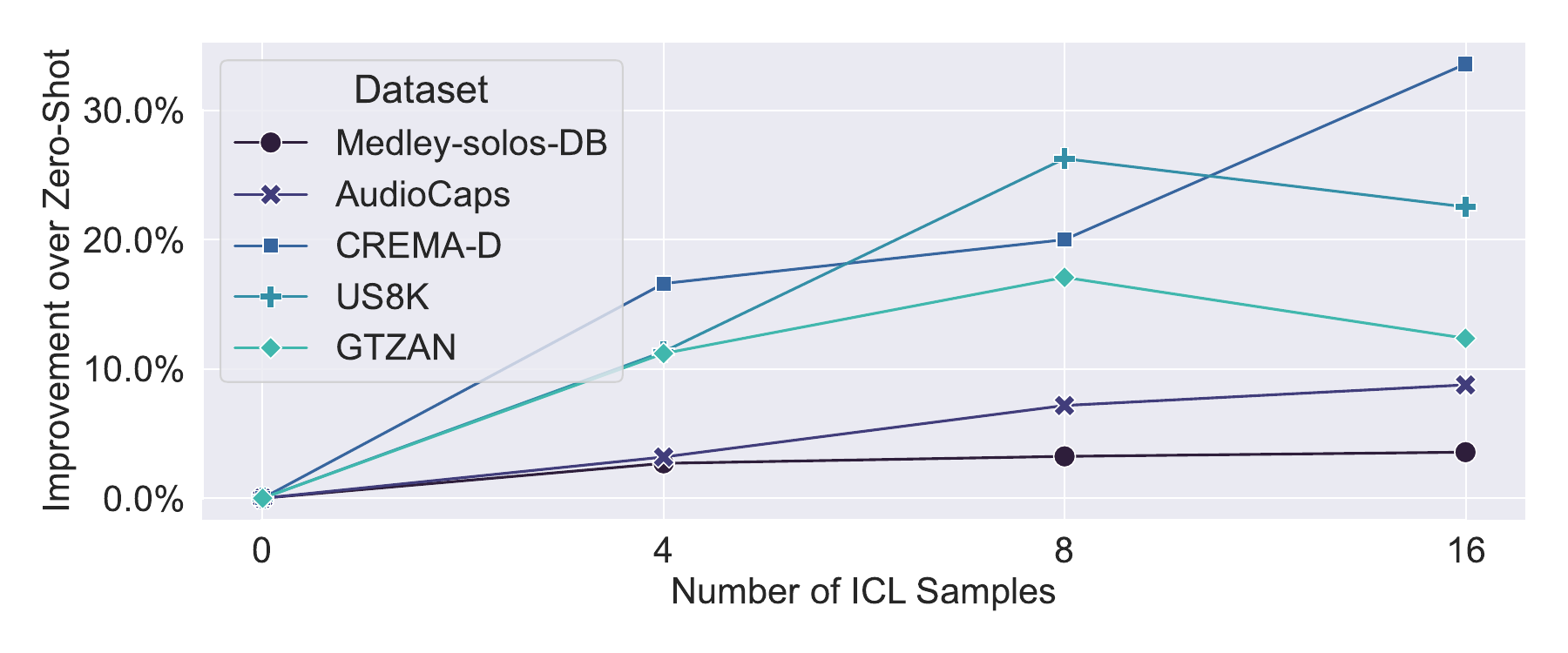}
    \vspace{-2.5em}
    \caption{\textit{Relative} improvement of few-shot results over zero-shot results under different number of ICL samples. Using more ICL samples consistently improves few-shot results, and the benefit is dataset-dependent.}
    \vspace{1em}
    \label{fig: ICL ablation}
\end{subfigure}
\begin{subfigure}
    \centering
    \includegraphics[width=0.48\textwidth]{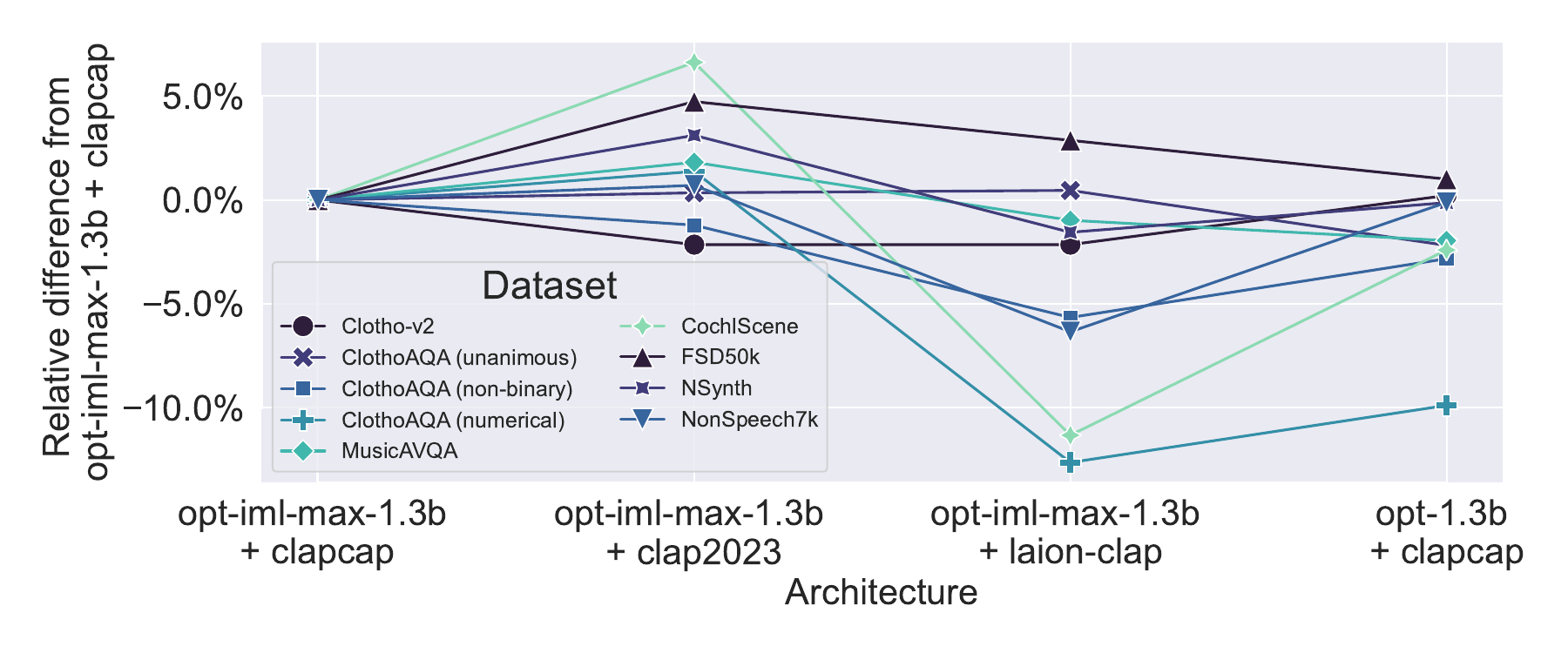}
    \vspace{-2.5em}
    \caption{\textit{Relative} difference of results from different LM backbones and audio encoders. The \texttt{opt-1.3b} model without instruction-tuning is systematically worse than the instruction-tuned \texttt{opt-iml-max-1.3b}. LAION-CLAP is worse than ClapCap in most cases. Clap2023 is better than ClapCap in close-ended tasks, but worse in open-ended tasks including captioning and open question-answering.}
    \vspace{1em}
    \label{fig: arch ablation}
\end{subfigure}
\begin{subfigure}
    \centering
    \includegraphics[width=0.48\textwidth]{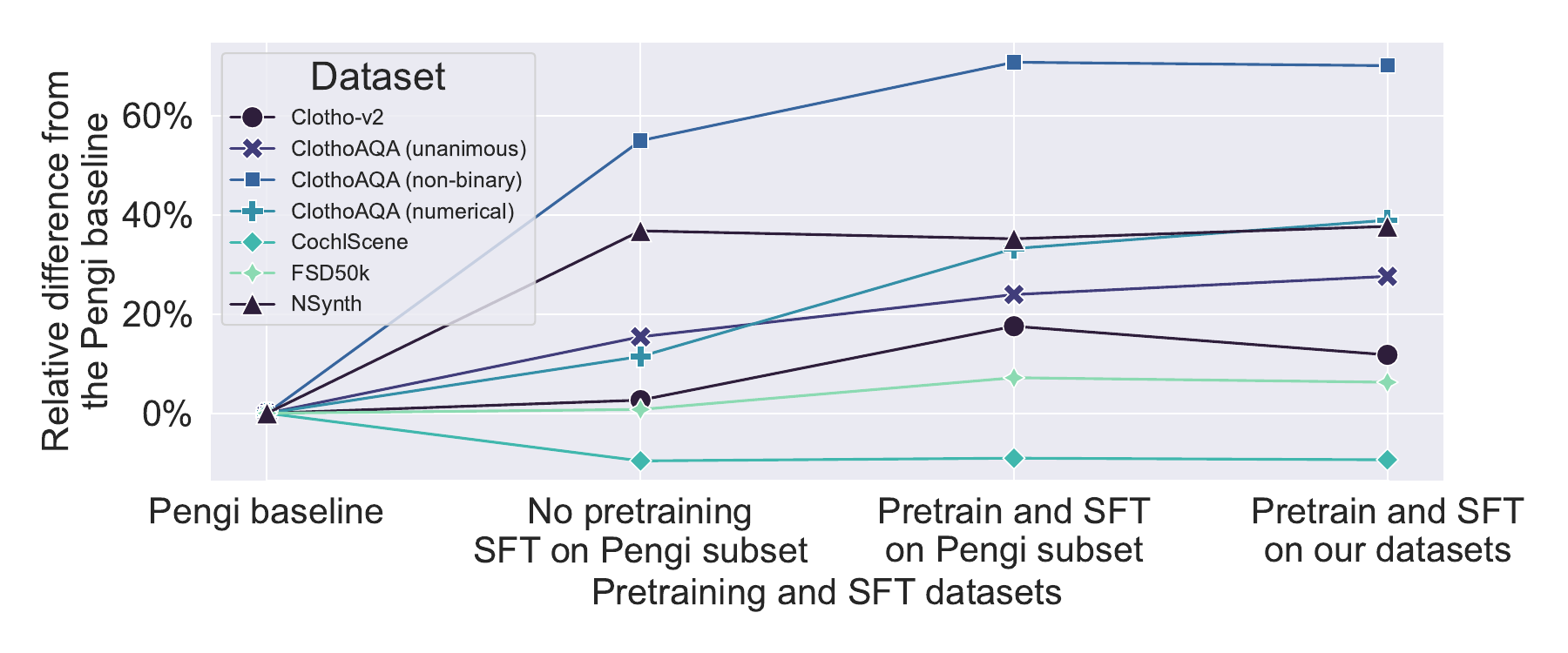}
    \vspace{-2.5em}
    \caption{\textit{Relative} difference of results from different training datasets compared to the Pengi baseline. Audio Flamingo achieves better results than Pengi even if no extra data is used. Larger scale pretraining and SFT lead to better results on these benchmarks on average.}
    \label{fig: data ablation}
\end{subfigure}
\end{figure}

\subsection{Q4: Ablation Studies}

\textbf{Effect of number of few-shot samples.}
We study different numbers of in-context few-shot samples and evaluate on the few-shot benchmarks. In Figure~\ref{fig: ICL ablation}, we plot the relative improvements over zero-shot results. Results show a clear trend that having more ICL samples improves few-shot results, and the improvement highly depends on the dataset.

\textbf{Effect of architecture.}
In Figure~\ref{fig: arch ablation}, we compare results between \texttt{opt-1.3b} \citep{zhang2022opt} and the instruction-tuned \texttt{opt-iml-max-1.3b} \citep{iyer2022opt}, and also compare different audio encoders including ClapCap \citep{CLAP2023}, Clap2023 \citep{CLAP2023}, and LAION-CLAP \citep{wu2023large}. In terms of the LM backbone, the instruction-tuned \texttt{opt} is better in most tasks. As of audio encoders, LAION-CLAP is worse in most tasks, ClapCap is better in open-ended tasks, and Clap2023 is better in most close-ended tasks.

\textbf{Effect of training data.} 
In Figure~\ref{fig: data ablation}, we compare Audio Flamingo trained on three different training sets to the Pengi baseline: (1) no pretraining, SFT on our best available dataset that is strictly a subset of Pengi's training set, (2) pretraining and SFT on this strict subset, and (3) pretraining and SFT on our curated datasets. Audio Flamingo achieves better evaluation results than Pengi even if no extra data is used. In addition, increasing the data amount in pretraining and SFT can improve the results on average.

\section{Conclusion and Future Work}\label{sec:discussion}
In this paper, we present Audio Flamingo, an audio language model with a series of innovations that achieves the state-of-the-art results on several close-ended and open-ended audio understanding tasks without task specific fine-tuning. It also has strong ICL and RAG abilities, and has the state-of-the-art few-shot learning results. Furthermore, we design a dataset generation strategy and introduce two dialogue datasets, enabling Audio Flamingo to chat with a user about the audio for multiple rounds and achieve state-of-the-art results on dialogue benchmarks.

Our model has several limitations, which we plan to address in future work. 
One important future direction is to investigate scaling strategies for using larger LMs. Assuming that larger LMs could have better knowledge and stronger ability to follow instructions, we believe that Audio Flamingo could benefit from a larger LM. 
A second future direction is to investigate complex speech-related tasks beyond transcription. This requires our model to condition on dense embeddings \citep{chen2023salm}. This modification is straightforward in Audio Flamingo as the architecture is flexible enough to support new embeddings through the addition of new cross-attention heads. 
Another future direction is to build a audio language model that can output both text and audio and follow more complex interleaved instructions. 
Finally, a future direction towards unifying more modalities is to combine the audio understanding abilities of our model with visual language models \citep{alayrac2022flamingo} such that one model could understand image, video, and audio.

\clearpage
\newpage
\section*{Acknowledgement}
We thank Siddharth Gururani, Zihan Liu, Mostofa Patwary, Shrimai Prabhumoye, and Chen Zhu for helpful discussions. We thank Ke Chen and Yuan Gong for help on sharing datasets with us.
\section*{Impact Statement}
This paper presents work whose goal is to advance the field of Machine Learning. The proposed method facilitates automation in the audio-language domain and may be used in a variety of scenarios such as education, healthcare, environment, industry, music, etc. Careful use of the model is essential to ensure compliance with copyright restrictions in certain situations.

\bibliographystyle{icml2024}
\bibliography{main}\label{sec:references}

\begin{thebibliography}{92}
\providecommand{\natexlab}[1]{#1}
\providecommand{\url}[1]{\texttt{#1}}
\expandafter\ifx\csname urlstyle\endcsname\relax
  \providecommand{\doi}[1]{doi: #1}\else
  \providecommand{\doi}{doi: \begingroup \urlstyle{rm}\Url}\fi

\bibitem[Achiam et~al.(2023)Achiam, Adler, Agarwal, Ahmad, Akkaya, Aleman,
  Almeida, Altenschmidt, Altman, Anadkat, et~al.]{achiam2023gpt}
Achiam, J., Adler, S., Agarwal, S., Ahmad, L., Akkaya, I., Aleman, F.~L.,
  Almeida, D., Altenschmidt, J., Altman, S., Anadkat, S., et~al.
\newblock Gpt-4 technical report.
\newblock \emph{arXiv preprint arXiv:2303.08774}, 2023.

\bibitem[Adigwe et~al.(2018)Adigwe, Tits, Haddad, Ostadabbas, and
  Dutoit]{adigwe2018emotional}
Adigwe, A., Tits, N., Haddad, K.~E., Ostadabbas, S., and Dutoit, T.
\newblock The emotional voices database: Towards controlling the emotion
  dimension in voice generation systems.
\newblock \emph{arXiv preprint arXiv:1806.09514}, 2018.

\bibitem[Agostinelli et~al.(2023)Agostinelli, Denk, Borsos, Engel, Verzetti,
  Caillon, Huang, Jansen, Roberts, Tagliasacchi,
  et~al.]{agostinelli2023musiclm}
Agostinelli, A., Denk, T.~I., Borsos, Z., Engel, J., Verzetti, M., Caillon, A.,
  Huang, Q., Jansen, A., Roberts, A., Tagliasacchi, M., et~al.
\newblock Musiclm: Generating music from text.
\newblock \emph{arXiv preprint arXiv:2301.11325}, 2023.

\bibitem[Alayrac et~al.(2022)Alayrac, Donahue, Luc, Miech, Barr, Hasson, Lenc,
  Mensch, Millican, Reynolds, et~al.]{alayrac2022flamingo}
Alayrac, J.-B., Donahue, J., Luc, P., Miech, A., Barr, I., Hasson, Y., Lenc,
  K., Mensch, A., Millican, K., Reynolds, M., et~al.
\newblock Flamingo: a visual language model for few-shot learning.
\newblock \emph{Advances in Neural Information Processing Systems},
  35:\penalty0 23716--23736, 2022.

\bibitem[Barros et~al.(2018)Barros, Churamani, Lakomkin, Siqueira, Sutherland,
  and Wermter]{barros2018omg}
Barros, P., Churamani, N., Lakomkin, E., Siqueira, H., Sutherland, A., and
  Wermter, S.
\newblock The omg-emotion behavior dataset.
\newblock In \emph{2018 International Joint Conference on Neural Networks
  (IJCNN)}, pp.\  1--7. IEEE, 2018.

\bibitem[Bogdanov et~al.(2019)Bogdanov, Won, Tovstogan, Porter, and
  Serra]{bogdanov2019mtg}
Bogdanov, D., Won, M., Tovstogan, P., Porter, A., and Serra, X.
\newblock The mtg-jamendo dataset for automatic music tagging.
\newblock ICML, 2019.

\bibitem[Borgeaud et~al.(2022)Borgeaud, Mensch, Hoffmann, Cai, Rutherford,
  Millican, Van Den~Driessche, Lespiau, Damoc, Clark,
  et~al.]{borgeaud2022improving}
Borgeaud, S., Mensch, A., Hoffmann, J., Cai, T., Rutherford, E., Millican, K.,
  Van Den~Driessche, G.~B., Lespiau, J.-B., Damoc, B., Clark, A., et~al.
\newblock Improving language models by retrieving from trillions of tokens.
\newblock In \emph{International conference on machine learning}, pp.\
  2206--2240. PMLR, 2022.

\bibitem[Brown et~al.(2020)Brown, Mann, Ryder, Subbiah, Kaplan, Dhariwal,
  Neelakantan, Shyam, Sastry, Askell, et~al.]{brown2020language}
Brown, T., Mann, B., Ryder, N., Subbiah, M., Kaplan, J.~D., Dhariwal, P.,
  Neelakantan, A., Shyam, P., Sastry, G., Askell, A., et~al.
\newblock Language models are few-shot learners.
\newblock \emph{Advances in neural information processing systems},
  33:\penalty0 1877--1901, 2020.

\bibitem[Cao et~al.(2014)Cao, Cooper, Keutmann, Gur, Nenkova, and
  Verma]{cao2014crema}
Cao, H., Cooper, D.~G., Keutmann, M.~K., Gur, R.~C., Nenkova, A., and Verma, R.
\newblock Crema-d: Crowd-sourced emotional multimodal actors dataset.
\newblock \emph{IEEE transactions on affective computing}, 5\penalty0
  (4):\penalty0 377--390, 2014.

\bibitem[Cartwright et~al.(2019)Cartwright, Mendez, Cramer, Lostanlen, Dove,
  Wu, Salamon, Nov, and Bello]{cartwright2019sonyc}
Cartwright, M., Mendez, A. E.~M., Cramer, A., Lostanlen, V., Dove, G., Wu,
  H.-H., Salamon, J., Nov, O., and Bello, J.
\newblock Sonyc urban sound tagging (sonyc-ust): A multilabel dataset from an
  urban acoustic sensor network.
\newblock 2019.

\bibitem[Chen et~al.(2022)Chen, Du, Zhu, Ma, Berg-Kirkpatrick, and
  Dubnov]{chen2022hts}
Chen, K., Du, X., Zhu, B., Ma, Z., Berg-Kirkpatrick, T., and Dubnov, S.
\newblock Hts-at: A hierarchical token-semantic audio transformer for sound
  classification and detection.
\newblock In \emph{ICASSP 2022-2022 IEEE International Conference on Acoustics,
  Speech and Signal Processing (ICASSP)}, pp.\  646--650. IEEE, 2022.

\bibitem[Chen et~al.(2023)Chen, Huang, Andrusenko, Hrinchuk, Puvvada, Li,
  Ghosh, Balam, and Ginsburg]{chen2023salm}
Chen, Z., Huang, H., Andrusenko, A., Hrinchuk, O., Puvvada, K.~C., Li, J.,
  Ghosh, S., Balam, J., and Ginsburg, B.
\newblock Salm: Speech-augmented language model with in-context learning for
  speech recognition and translation.
\newblock \emph{arXiv preprint arXiv:2310.09424}, 2023.

\bibitem[Chu et~al.(2023)Chu, Xu, Zhou, Yang, Zhang, Yan, Zhou, and
  Zhou]{chu2023qwen}
Chu, Y., Xu, J., Zhou, X., Yang, Q., Zhang, S., Yan, Z., Zhou, C., and Zhou, J.
\newblock Qwen-audio: Advancing universal audio understanding via unified
  large-scale audio-language models.
\newblock \emph{arXiv preprint arXiv:2311.07919}, 2023.

\bibitem[Defferrard et~al.(2016)Defferrard, Benzi, Vandergheynst, and
  Bresson]{defferrard2016fma}
Defferrard, M., Benzi, K., Vandergheynst, P., and Bresson, X.
\newblock Fma: A dataset for music analysis.
\newblock \emph{arXiv preprint arXiv:1612.01840}, 2016.

\bibitem[D{\'e}fossez et~al.(2022)D{\'e}fossez, Copet, Synnaeve, and
  Adi]{defossez2022high}
D{\'e}fossez, A., Copet, J., Synnaeve, G., and Adi, Y.
\newblock High fidelity neural audio compression.
\newblock \emph{arXiv preprint arXiv:2210.13438}, 2022.

\bibitem[Deshmukh et~al.(2022)Deshmukh, Elizalde, and Wang]{deshmukh2022audio}
Deshmukh, S., Elizalde, B., and Wang, H.
\newblock Audio retrieval with wavtext5k and clap training.
\newblock \emph{arXiv preprint arXiv:2209.14275}, 2022.

\bibitem[Deshmukh et~al.(2023)Deshmukh, Elizalde, Singh, and
  Wang]{deshmukh2023pengi}
Deshmukh, S., Elizalde, B., Singh, R., and Wang, H.
\newblock Pengi: An audio language model for audio tasks.
\newblock \emph{arXiv preprint arXiv:2305.11834}, 2023.

\bibitem[Doh et~al.(2023)Doh, Choi, Lee, and Nam]{doh2023lp}
Doh, S., Choi, K., Lee, J., and Nam, J.
\newblock Lp-musiccaps: Llm-based pseudo music captioning.
\newblock \emph{arXiv preprint arXiv:2307.16372}, 2023.

\bibitem[Driess et~al.(2023)Driess, Xia, Sajjadi, Lynch, Chowdhery, Ichter,
  Wahid, Tompson, Vuong, Yu, et~al.]{driess2023palm}
Driess, D., Xia, F., Sajjadi, M.~S., Lynch, C., Chowdhery, A., Ichter, B.,
  Wahid, A., Tompson, J., Vuong, Q., Yu, T., et~al.
\newblock Palm-e: An embodied multimodal language model.
\newblock \emph{arXiv preprint arXiv:2303.03378}, 2023.

\bibitem[Drossos et~al.(2020)Drossos, Lipping, and Virtanen]{drossos2020clotho}
Drossos, K., Lipping, S., and Virtanen, T.
\newblock Clotho: An audio captioning dataset.
\newblock In \emph{ICASSP 2020-2020 IEEE International Conference on Acoustics,
  Speech and Signal Processing (ICASSP)}, pp.\  736--740. IEEE, 2020.

\bibitem[Duan et~al.(2023)Duan, Wei, Wang, Liu, Fang, Zhang, Lin, and
  Chen]{duan2023botchat}
Duan, H., Wei, J., Wang, C., Liu, H., Fang, Y., Zhang, S., Lin, D., and Chen,
  K.
\newblock Botchat: Evaluating llms' capabilities of having multi-turn
  dialogues.
\newblock \emph{arXiv preprint arXiv:2310.13650}, 2023.

\bibitem[Elizalde et~al.(2023{\natexlab{a}})Elizalde, Deshmukh, Al~Ismail, and
  Wang]{CLAP2022}
Elizalde, B., Deshmukh, S., Al~Ismail, M., and Wang, H.
\newblock Clap learning audio concepts from natural language supervision.
\newblock In \emph{ICASSP 2023-2023 IEEE International Conference on Acoustics,
  Speech and Signal Processing (ICASSP)}, pp.\  1--5. IEEE, 2023{\natexlab{a}}.

\bibitem[Elizalde et~al.(2023{\natexlab{b}})Elizalde, Deshmukh, and
  Wang]{CLAP2023}
Elizalde, B., Deshmukh, S., and Wang, H.
\newblock Natural language supervision for general-purpose audio
  representations, 2023{\natexlab{b}}.
\newblock URL \url{https://arxiv.org/abs/2309.05767}.

\bibitem[Engel et~al.(2017)Engel, Resnick, Roberts, Dieleman, Eck, Simonyan,
  and Norouzi]{nsynth2017}
Engel, J., Resnick, C., Roberts, A., Dieleman, S., Eck, D., Simonyan, K., and
  Norouzi, M.
\newblock Neural audio synthesis of musical notes with wavenet autoencoders,
  2017.

\bibitem[Fonseca et~al.(2021)Fonseca, Favory, Pons, Font, and
  Serra]{fonseca2021fsd50k}
Fonseca, E., Favory, X., Pons, J., Font, F., and Serra, X.
\newblock Fsd50k: an open dataset of human-labeled sound events.
\newblock \emph{IEEE/ACM Transactions on Audio, Speech, and Language
  Processing}, 30:\penalty0 829--852, 2021.

\bibitem[Foster et~al.(2015)Foster, Sigtia, Krstulovic, Barker, and
  Plumbley]{foster2015chime}
Foster, P., Sigtia, S., Krstulovic, S., Barker, J., and Plumbley, M.~D.
\newblock Chime-home: A dataset for sound source recognition in a domestic
  environment.
\newblock In \emph{2015 IEEE Workshop on Applications of Signal Processing to
  Audio and Acoustics (WASPAA)}, pp.\  1--5. IEEE, 2015.

\bibitem[Gao et~al.(2022)Gao, Ni, Qian, Zhang, Chang, and
  Hasegawa-Johnson]{gao2022wavprompt}
Gao, H., Ni, J., Qian, K., Zhang, Y., Chang, S., and Hasegawa-Johnson, M.
\newblock Wavprompt: Towards few-shot spoken language understanding with frozen
  language models.
\newblock \emph{arXiv preprint arXiv:2203.15863}, 2022.

\bibitem[Gardner et~al.(2023)Gardner, Durand, Stoller, and
  Bittner]{gardner2023llark}
Gardner, J., Durand, S., Stoller, D., and Bittner, R.~M.
\newblock Llark: A multimodal foundation model for music.
\newblock \emph{arXiv preprint arXiv:2310.07160}, 2023.

\bibitem[Gemmeke et~al.(2017)Gemmeke, Ellis, Freedman, Jansen, Lawrence, Moore,
  Plakal, and Ritter]{gemmeke2017audio}
Gemmeke, J.~F., Ellis, D.~P., Freedman, D., Jansen, A., Lawrence, W., Moore,
  R.~C., Plakal, M., and Ritter, M.
\newblock Audio set: An ontology and human-labeled dataset for audio events.
\newblock In \emph{2017 IEEE international conference on acoustics, speech and
  signal processing (ICASSP)}, pp.\  776--780. IEEE, 2017.

\bibitem[Ghosh et~al.(2023)Ghosh, Kumar, Evuru, Duraiswami, and
  Manocha]{ghosh2023recap}
Ghosh, S., Kumar, S., Evuru, C. K.~R., Duraiswami, R., and Manocha, D.
\newblock Recap: Retrieval-augmented audio captioning.
\newblock \emph{arXiv preprint arXiv:2309.09836}, 2023.

\bibitem[Gong et~al.(2021)Gong, Chung, and Glass]{gong2021ast}
Gong, Y., Chung, Y.-A., and Glass, J.
\newblock Ast: Audio spectrogram transformer.
\newblock \emph{arXiv preprint arXiv:2104.01778}, 2021.

\bibitem[Gong et~al.(2023{\natexlab{a}})Gong, Khurana, Karlinsky, and
  Glass]{gong2023whisper}
Gong, Y., Khurana, S., Karlinsky, L., and Glass, J.
\newblock Whisper-at: Noise-robust automatic speech recognizers are also strong
  general audio event taggers.
\newblock \emph{arXiv preprint arXiv:2307.03183}, 2023{\natexlab{a}}.

\bibitem[Gong et~al.(2023{\natexlab{b}})Gong, Liu, Luo, Karlinsky, and
  Glass]{gong2023joint}
Gong, Y., Liu, A., Luo, H., Karlinsky, L., and Glass, J.
\newblock Joint audio and speech understanding.
\newblock In \emph{IEEE Automatic Speech Recognition and Understanding
  Workshop}, 2023{\natexlab{b}}.

\bibitem[Gong et~al.(2023{\natexlab{c}})Gong, Luo, Liu, Karlinsky, and
  Glass]{gong2023ltu}
Gong, Y., Luo, H., Liu, A.~H., Karlinsky, L., and Glass, J.
\newblock Listen, think, and understand.
\newblock \emph{arXiv preprint arXiv:2305.10790}, 2023{\natexlab{c}}.

\bibitem[Guu et~al.(2020)Guu, Lee, Tung, Pasupat, and Chang]{guu2020retrieval}
Guu, K., Lee, K., Tung, Z., Pasupat, P., and Chang, M.
\newblock Retrieval augmented language model pre-training.
\newblock In \emph{International conference on machine learning}, pp.\
  3929--3938. PMLR, 2020.

\bibitem[Han et~al.(2023)Han, Zhang, Shao, Gao, Xu, Xiao, Zhang, Liu, Wen, Guo,
  et~al.]{han2023imagebind}
Han, J., Zhang, R., Shao, W., Gao, P., Xu, P., Xiao, H., Zhang, K., Liu, C.,
  Wen, S., Guo, Z., et~al.
\newblock Imagebind-llm: Multi-modality instruction tuning.
\newblock \emph{arXiv preprint arXiv:2309.03905}, 2023.

\bibitem[Hershey et~al.(2021)Hershey, Ellis, Fonseca, Jansen, Liu, Moore, and
  Plakal]{hershey2021benefit}
Hershey, S., Ellis, D.~P., Fonseca, E., Jansen, A., Liu, C., Moore, R.~C., and
  Plakal, M.
\newblock The benefit of temporally-strong labels in audio event
  classification.
\newblock In \emph{ICASSP 2021-2021 IEEE International Conference on Acoustics,
  Speech and Signal Processing (ICASSP)}, pp.\  366--370. IEEE, 2021.

\bibitem[Hsu et~al.(2023)Hsu, Chang, Li, and Lee]{hsu2023exploration}
Hsu, M.-H., Chang, K.-W., Li, S.-W., and Lee, H.-y.
\newblock An exploration of in-context learning for speech language model.
\newblock \emph{arXiv preprint arXiv:2310.12477}, 2023.

\bibitem[Huang et~al.(2022)Huang, Jansen, Lee, Ganti, Li, and
  Ellis]{huang2022mulan}
Huang, Q., Jansen, A., Lee, J., Ganti, R., Li, J.~Y., and Ellis, D.~P.
\newblock Mulan: A joint embedding of music audio and natural language.
\newblock \emph{arXiv preprint arXiv:2208.12415}, 2022.

\bibitem[Huang et~al.(2023)Huang, Li, Yang, Shi, Chang, Ye, Wu, Hong, Huang,
  Liu, et~al.]{huang2023audiogpt}
Huang, R., Li, M., Yang, D., Shi, J., Chang, X., Ye, Z., Wu, Y., Hong, Z.,
  Huang, J., Liu, J., et~al.
\newblock Audiogpt: Understanding and generating speech, music, sound, and
  talking head.
\newblock \emph{arXiv preprint arXiv:2304.12995}, 2023.

\bibitem[Iyer et~al.(2022)Iyer, Lin, Pasunuru, Mihaylov, Simig, Yu, Shuster,
  Wang, Liu, Koura, et~al.]{iyer2022opt}
Iyer, S., Lin, X.~V., Pasunuru, R., Mihaylov, T., Simig, D., Yu, P., Shuster,
  K., Wang, T., Liu, Q., Koura, P.~S., et~al.
\newblock Opt-iml: Scaling language model instruction meta learning through the
  lens of generalization.
\newblock \emph{arXiv preprint arXiv:2212.12017}, 2022.

\bibitem[James et~al.(2018)James, Tian, and Watson]{james2018open}
James, J., Tian, L., and Watson, C.
\newblock An open source emotional speech corpus for human robot interaction
  applications.
\newblock \emph{Interspeech 2018}, 2018.

\bibitem[Jeong \& Park(2022)Jeong and Park]{jeong2022cochlscene}
Jeong, I.-Y. and Park, J.
\newblock Cochlscene: Acquisition of acoustic scene data using crowdsourcing.
\newblock In \emph{2022 Asia-Pacific Signal and Information Processing
  Association Annual Summit and Conference (APSIPA ASC)}, pp.\  17--21. IEEE,
  2022.

\bibitem[Johnson et~al.(2019)Johnson, Douze, and J{\'e}gou]{johnson2019billion}
Johnson, J., Douze, M., and J{\'e}gou, H.
\newblock Billion-scale similarity search with {GPUs}.
\newblock \emph{IEEE Transactions on Big Data}, 7\penalty0 (3):\penalty0
  535--547, 2019.

\bibitem[Karpukhin et~al.(2020)Karpukhin, O{\u{g}}uz, Min, Lewis, Wu, Edunov,
  Chen, and Yih]{karpukhin2020dense}
Karpukhin, V., O{\u{g}}uz, B., Min, S., Lewis, P., Wu, L., Edunov, S., Chen,
  D., and Yih, W.-t.
\newblock Dense passage retrieval for open-domain question answering.
\newblock \emph{arXiv preprint arXiv:2004.04906}, 2020.

\bibitem[Khomenko et~al.(2016)Khomenko, Shyshkov, Radyvonenko, and
  Bokhan]{khomenko2016accelerating}
Khomenko, V., Shyshkov, O., Radyvonenko, O., and Bokhan, K.
\newblock Accelerating recurrent neural network training using sequence
  bucketing and multi-gpu data parallelization.
\newblock In \emph{2016 IEEE First International Conference on Data Stream
  Mining \& Processing (DSMP)}, pp.\  100--103. IEEE, 2016.

\bibitem[Kim et~al.(2019)Kim, Kim, Lee, and Kim]{kim2019audiocaps}
Kim, C.~D., Kim, B., Lee, H., and Kim, G.
\newblock Audiocaps: Generating captions for audios in the wild.
\newblock In \emph{Proceedings of the 2019 Conference of the North American
  Chapter of the Association for Computational Linguistics: Human Language
  Technologies, Volume 1 (Long and Short Papers)}, pp.\  119--132, 2019.

\bibitem[Kong et~al.(2020)Kong, Cao, Iqbal, Wang, Wang, and
  Plumbley]{kong2020panns}
Kong, Q., Cao, Y., Iqbal, T., Wang, Y., Wang, W., and Plumbley, M.~D.
\newblock Panns: Large-scale pretrained audio neural networks for audio pattern
  recognition.
\newblock \emph{IEEE/ACM Transactions on Audio, Speech, and Language
  Processing}, 28:\penalty0 2880--2894, 2020.

\bibitem[Lewis et~al.(2020)Lewis, Perez, Piktus, Petroni, Karpukhin, Goyal,
  K{\"u}ttler, Lewis, Yih, Rockt{\"a}schel, et~al.]{lewis2020retrieval}
Lewis, P., Perez, E., Piktus, A., Petroni, F., Karpukhin, V., Goyal, N.,
  K{\"u}ttler, H., Lewis, M., Yih, W.-t., Rockt{\"a}schel, T., et~al.
\newblock Retrieval-augmented generation for knowledge-intensive nlp tasks.
\newblock \emph{Advances in Neural Information Processing Systems},
  33:\penalty0 9459--9474, 2020.

\bibitem[Li et~al.(2022)Li, Wei, Tian, Xu, Wen, and Hu]{li2022learning}
Li, G., Wei, Y., Tian, Y., Xu, C., Wen, J.-R., and Hu, D.
\newblock Learning to answer questions in dynamic audio-visual scenarios.
\newblock In \emph{Proceedings of the IEEE/CVF Conference on Computer Vision
  and Pattern Recognition}, pp.\  19108--19118, 2022.

\bibitem[Li et~al.(2023{\natexlab{a}})Li, Li, Savarese, and Hoi]{li2023blip}
Li, J., Li, D., Savarese, S., and Hoi, S.
\newblock Blip-2: Bootstrapping language-image pre-training with frozen image
  encoders and large language models.
\newblock \emph{arXiv preprint arXiv:2301.12597}, 2023{\natexlab{a}}.

\bibitem[Li et~al.(2023{\natexlab{b}})Li, Yuan, Zhang, Ma, Chen, Yin, Lin,
  Ragni, Benetos, Gyenge, et~al.]{li2023mert}
Li, Y., Yuan, R., Zhang, G., Ma, Y., Chen, X., Yin, H., Lin, C., Ragni, A.,
  Benetos, E., Gyenge, N., et~al.
\newblock Mert: Acoustic music understanding model with large-scale
  self-supervised training.
\newblock \emph{arXiv preprint arXiv:2306.00107}, 2023{\natexlab{b}}.

\bibitem[Lin(2004)]{lin2004rouge}
Lin, C.-Y.
\newblock Rouge: A package for automatic evaluation of summaries.
\newblock In \emph{Text summarization branches out}, pp.\  74--81, 2004.

\bibitem[Lipping et~al.(2022)Lipping, Sudarsanam, Drossos, and
  Virtanen]{lipping2022clotho}
Lipping, S., Sudarsanam, P., Drossos, K., and Virtanen, T.
\newblock Clotho-aqa: A crowdsourced dataset for audio question answering.
\newblock In \emph{2022 30th European Signal Processing Conference (EUSIPCO)},
  pp.\  1140--1144. IEEE, 2022.

\bibitem[Liu et~al.(2023{\natexlab{a}})Liu, Li, Wu, and Lee]{liu2023visual}
Liu, H., Li, C., Wu, Q., and Lee, Y.~J.
\newblock Visual instruction tuning.
\newblock \emph{arXiv preprint arXiv:2304.08485}, 2023{\natexlab{a}}.

\bibitem[Liu et~al.(2023{\natexlab{b}})Liu, Hussain, Sun, and
  Shan]{liu2023music}
Liu, S., Hussain, A.~S., Sun, C., and Shan, Y.
\newblock Music understanding llama: Advancing text-to-music generation with
  question answering and captioning.
\newblock \emph{arXiv preprint arXiv:2308.11276}, 2023{\natexlab{b}}.

\bibitem[Livingstone \& Russo(2018)Livingstone and
  Russo]{livingstone2018ryerson}
Livingstone, S.~R. and Russo, F.~A.
\newblock The ryerson audio-visual database of emotional speech and song
  (ravdess): A dynamic, multimodal set of facial and vocal expressions in north
  american english.
\newblock \emph{PloS one}, 13\penalty0 (5):\penalty0 e0196391, 2018.

\bibitem[Loshchilov \& Hutter(2017)Loshchilov and
  Hutter]{loshchilov2017decoupled}
Loshchilov, I. and Hutter, F.
\newblock Decoupled weight decay regularization.
\newblock \emph{arXiv preprint arXiv:1711.05101}, 2017.

\bibitem[Lostanlen et~al.(2019)Lostanlen, Cella, Bittner, and
  Essid]{lostanlen_2019_1344103}
Lostanlen, V., Cella, C.-E., Bittner, R., and Essid, S.
\newblock {Medley-solos-DB: a cross-collection dataset for musical instrument
  recognition}, February 2019.
\newblock URL \url{https://doi.org/10.5281/zenodo.1344103}.

\bibitem[Lotfian \& Busso(2017)Lotfian and Busso]{lotfian2017building}
Lotfian, R. and Busso, C.
\newblock Building naturalistic emotionally balanced speech corpus by
  retrieving emotional speech from existing podcast recordings.
\newblock \emph{IEEE Transactions on Affective Computing}, 10\penalty0
  (4):\penalty0 471--483, 2017.

\bibitem[Lyu et~al.(2023)Lyu, Wu, Wang, Huang, Liu, Du, Shi, and
  Tu]{lyu2023macaw}
Lyu, C., Wu, M., Wang, L., Huang, X., Liu, B., Du, Z., Shi, S., and Tu, Z.
\newblock Macaw-llm: Multi-modal language modeling with image, audio, video,
  and text integration.
\newblock \emph{arXiv preprint arXiv:2306.09093}, 2023.

\bibitem[Martin~Morato \& Mesaros(2021)Martin~Morato and
  Mesaros]{martin2021diversity}
Martin~Morato, I. and Mesaros, A.
\newblock Diversity and bias in audio captioning datasets.
\newblock 2021.

\bibitem[Mei et~al.(2023)Mei, Meng, Liu, Kong, Ko, Zhao, Plumbley, Zou, and
  Wang]{mei2023wavcaps}
Mei, X., Meng, C., Liu, H., Kong, Q., Ko, T., Zhao, C., Plumbley, M.~D., Zou,
  Y., and Wang, W.
\newblock Wavcaps: A chatgpt-assisted weakly-labelled audio captioning dataset
  for audio-language multimodal research.
\newblock \emph{arXiv preprint arXiv:2303.17395}, 2023.

\bibitem[Mohanty()]{birds2022}
Mohanty, S.~P.
\newblock {Sound Of 114 Species Of Birds Till 2022}.
\newblock URL
  \url{https://www.kaggle.com/datasets/soumendraprasad/sound-of-114-species-of-birds-till-2022}.

\bibitem[Moon et~al.(2023)Moon, Madotto, Lin, Nagarajan, Smith, Jain, Yeh,
  Murugesan, Heidari, Liu, et~al.]{moon2023anymal}
Moon, S., Madotto, A., Lin, Z., Nagarajan, T., Smith, M., Jain, S., Yeh, C.-F.,
  Murugesan, P., Heidari, P., Liu, Y., et~al.
\newblock Anymal: An efficient and scalable any-modality augmented language
  model.
\newblock \emph{arXiv preprint arXiv:2309.16058}, 2023.

\bibitem[Oncescu et~al.(2021)Oncescu, Koepke, Henriques, Akata, and
  Albanie]{oncescu2021audio}
Oncescu, A.-M., Koepke, A., Henriques, J.~F., Akata, Z., and Albanie, S.
\newblock Audio retrieval with natural language queries.
\newblock \emph{arXiv preprint arXiv:2105.02192}, 2021.

\bibitem[Ouyang et~al.(2022)Ouyang, Wu, Jiang, Almeida, Wainwright, Mishkin,
  Zhang, Agarwal, Slama, Ray, et~al.]{ouyang2022training}
Ouyang, L., Wu, J., Jiang, X., Almeida, D., Wainwright, C., Mishkin, P., Zhang,
  C., Agarwal, S., Slama, K., Ray, A., et~al.
\newblock Training language models to follow instructions with human feedback.
\newblock \emph{Advances in Neural Information Processing Systems},
  35:\penalty0 27730--27744, 2022.

\bibitem[Papineni et~al.(2002)Papineni, Roukos, Ward, and
  Zhu]{papineni2002bleu}
Papineni, K., Roukos, S., Ward, T., and Zhu, W.-J.
\newblock Bleu: a method for automatic evaluation of machine translation.
\newblock In \emph{Proceedings of the 40th annual meeting of the Association
  for Computational Linguistics}, pp.\  311--318, 2002.

\bibitem[Park et~al.(2022)Park, Cho, Sim, Lee, and Choo]{park2022enemy}
Park, J., Cho, Y., Sim, G., Lee, H., and Choo, J.
\newblock Enemy spotted: In-game gun sound dataset for gunshot classification
  and localization.
\newblock In \emph{2022 IEEE Conference on Games (CoG)}, pp.\  56--63. IEEE,
  2022.

\bibitem[Pichora-Fuller \& Dupuis(2020)Pichora-Fuller and Dupuis]{tess}
Pichora-Fuller, M.~K. and Dupuis, K.
\newblock {Toronto emotional speech set (TESS)}, 2020.
\newblock URL \url{https://doi.org/10.5683/SP2/E8H2MF}.

\bibitem[Poria et~al.(2018)Poria, Hazarika, Majumder, Naik, Cambria, and
  Mihalcea]{poria2018meld}
Poria, S., Hazarika, D., Majumder, N., Naik, G., Cambria, E., and Mihalcea, R.
\newblock Meld: A multimodal multi-party dataset for emotion recognition in
  conversations.
\newblock \emph{arXiv preprint arXiv:1810.02508}, 2018.

\bibitem[Radford et~al.(2023)Radford, Kim, Xu, Brockman, McLeavey, and
  Sutskever]{radford2023robust}
Radford, A., Kim, J.~W., Xu, T., Brockman, G., McLeavey, C., and Sutskever, I.
\newblock Robust speech recognition via large-scale weak supervision.
\newblock In \emph{International Conference on Machine Learning}, pp.\
  28492--28518. PMLR, 2023.

\bibitem[Rafii et~al.(2019)Rafii, Liutkus, Stöter, Mimilakis, and
  Bittner]{musdb18-hq}
Rafii, Z., Liutkus, A., Stöter, F.-R., Mimilakis, S.~I., and Bittner, R.
\newblock Musdb18-hq - an uncompressed version of musdb18, August 2019.
\newblock URL \url{https://doi.org/10.5281/zenodo.3338373}.

\bibitem[Rashid et~al.(2023)Rashid, Li, and Du]{rashid2023nonspeech7k}
Rashid, M.~M., Li, G., and Du, C.
\newblock Nonspeech7k dataset: Classification and analysis of human non-speech
  sound.
\newblock \emph{IET Signal Processing}, 17\penalty0 (6):\penalty0 e12233, 2023.

\bibitem[Reimers \& Gurevych(2019)Reimers and
  Gurevych]{reimers-2019-sentence-bert}
Reimers, N. and Gurevych, I.
\newblock Sentence-bert: Sentence embeddings using siamese bert-networks.
\newblock In \emph{Proceedings of the 2019 Conference on Empirical Methods in
  Natural Language Processing}. Association for Computational Linguistics, 11
  2019.
\newblock URL \url{https://arxiv.org/abs/1908.10084}.

\bibitem[Reimers \& Gurevych(2020)Reimers and
  Gurevych]{reimers-2020-multilingual-sentence-bert}
Reimers, N. and Gurevych, I.
\newblock Making monolingual sentence embeddings multilingual using knowledge
  distillation.
\newblock In \emph{Proceedings of the 2020 Conference on Empirical Methods in
  Natural Language Processing}. Association for Computational Linguistics, 11
  2020.
\newblock URL \url{https://arxiv.org/abs/2004.09813}.

\bibitem[Rubenstein et~al.(2023)Rubenstein, Asawaroengchai, Nguyen, Bapna,
  Borsos, Quitry, Chen, Badawy, Han, Kharitonov,
  et~al.]{rubenstein2023audiopalm}
Rubenstein, P.~K., Asawaroengchai, C., Nguyen, D.~D., Bapna, A., Borsos, Z.,
  Quitry, F. d.~C., Chen, P., Badawy, D.~E., Han, W., Kharitonov, E., et~al.
\newblock Audiopalm: A large language model that can speak and listen.
\newblock \emph{arXiv preprint arXiv:2306.12925}, 2023.

\bibitem[Salamon et~al.(2014)Salamon, Jacoby, and Bello]{salamon2014dataset}
Salamon, J., Jacoby, C., and Bello, J.~P.
\newblock A dataset and taxonomy for urban sound research.
\newblock In \emph{Proceedings of the 22nd ACM international conference on
  Multimedia}, pp.\  1041--1044, 2014.

\bibitem[Salewski et~al.(2023)Salewski, Fauth, Koepke, and
  Akata]{salewski2023zero}
Salewski, L., Fauth, S., Koepke, A., and Akata, Z.
\newblock Zero-shot audio captioning with audio-language model guidance and
  audio context keywords.
\newblock \emph{arXiv preprint arXiv:2311.08396}, 2023.

\bibitem[Sturm(2013)]{sturm2013gtzan}
Sturm, B.~L.
\newblock The gtzan dataset: Its contents, its faults, their effects on
  evaluation, and its future use.
\newblock \emph{arXiv preprint arXiv:1306.1461}, 2013.

\bibitem[Tang et~al.(2023{\natexlab{a}})Tang, Yu, Sun, Chen, Tan, Li, Lu, Ma,
  and Zhang]{tang2023salmonn}
Tang, C., Yu, W., Sun, G., Chen, X., Tan, T., Li, W., Lu, L., Ma, Z., and
  Zhang, C.
\newblock Salmonn: Towards generic hearing abilities for large language models.
\newblock \emph{arXiv preprint arXiv:2310.13289}, 2023{\natexlab{a}}.

\bibitem[Tang et~al.(2023{\natexlab{b}})Tang, Yang, Khademi, Liu, Zhu, and
  Bansal]{tang2023codi}
Tang, Z., Yang, Z., Khademi, M., Liu, Y., Zhu, C., and Bansal, M.
\newblock Codi-2: In-context, interleaved, and interactive any-to-any
  generation.
\newblock \emph{arXiv preprint arXiv:2311.18775}, 2023{\natexlab{b}}.

\bibitem[Tsimpoukelli et~al.(2021)Tsimpoukelli, Menick, Cabi, Eslami, Vinyals,
  and Hill]{tsimpoukelli2021multimodal}
Tsimpoukelli, M., Menick, J.~L., Cabi, S., Eslami, S., Vinyals, O., and Hill,
  F.
\newblock Multimodal few-shot learning with frozen language models.
\newblock \emph{Advances in Neural Information Processing Systems},
  34:\penalty0 200--212, 2021.

\bibitem[Vaswani et~al.(2017)Vaswani, Shazeer, Parmar, Uszkoreit, Jones, Gomez,
  Kaiser, and Polosukhin]{vaswani2017attention}
Vaswani, A., Shazeer, N., Parmar, N., Uszkoreit, J., Jones, L., Gomez, A.~N.,
  Kaiser, {\L}., and Polosukhin, I.
\newblock Attention is all you need.
\newblock \emph{Advances in neural information processing systems}, 30, 2017.

\bibitem[Vedantam et~al.(2015)Vedantam, Lawrence~Zitnick, and
  Parikh]{vedantam2015cider}
Vedantam, R., Lawrence~Zitnick, C., and Parikh, D.
\newblock Cider: Consensus-based image description evaluation.
\newblock In \emph{Proceedings of the IEEE conference on computer vision and
  pattern recognition}, pp.\  4566--4575, 2015.

\bibitem[Wang et~al.(2023)Wang, Yang, Wu, and Zhang]{wang2023can}
Wang, S., Yang, C.-H.~H., Wu, J., and Zhang, C.
\newblock Can whisper perform speech-based in-context learning.
\newblock \emph{arXiv preprint arXiv:2309.07081}, 2023.

\bibitem[Wei et~al.(2021)Wei, Bosma, Zhao, Guu, Yu, Lester, Du, Dai, and
  Le]{wei2021finetuned}
Wei, J., Bosma, M., Zhao, V.~Y., Guu, K., Yu, A.~W., Lester, B., Du, N., Dai,
  A.~M., and Le, Q.~V.
\newblock Finetuned language models are zero-shot learners.
\newblock \emph{arXiv preprint arXiv:2109.01652}, 2021.

\bibitem[Won et~al.(2023)Won, Hung, and Le]{won2023foundation}
Won, M., Hung, Y.-N., and Le, D.
\newblock A foundation model for music informatics.
\newblock \emph{arXiv preprint arXiv:2311.03318}, 2023.

\bibitem[Wu et~al.(2023)Wu, Chen, Zhang, Hui, Berg-Kirkpatrick, and
  Dubnov]{wu2023large}
Wu, Y., Chen, K., Zhang, T., Hui, Y., Berg-Kirkpatrick, T., and Dubnov, S.
\newblock Large-scale contrastive language-audio pretraining with feature
  fusion and keyword-to-caption augmentation.
\newblock In \emph{ICASSP 2023-2023 IEEE International Conference on Acoustics,
  Speech and Signal Processing (ICASSP)}, pp.\  1--5. IEEE, 2023.

\bibitem[Yang et~al.(2023)Yang, Ping, Liu, Korthikanti, Nie, Huang, Fan, Yu,
  Lan, Li, et~al.]{yang2023re}
Yang, Z., Ping, W., Liu, Z., Korthikanti, V., Nie, W., Huang, D.-A., Fan, L.,
  Yu, Z., Lan, S., Li, B., et~al.
\newblock Re-vilm: Retrieval-augmented visual language model for zero and
  few-shot image captioning.
\newblock In \emph{EMNLP}, 2023.

\bibitem[Zhang et~al.(2022)Zhang, Roller, Goyal, Artetxe, Chen, Chen, Dewan,
  Diab, Li, Lin, et~al.]{zhang2022opt}
Zhang, S., Roller, S., Goyal, N., Artetxe, M., Chen, M., Chen, S., Dewan, C.,
  Diab, M., Li, X., Lin, X.~V., et~al.
\newblock Opt: Open pre-trained transformer language models.
\newblock \emph{arXiv preprint arXiv:2205.01068}, 2022.

\bibitem[Zhao et~al.(2023)Zhao, Guo, Yue, Chen, Shao, Zhu, Yuan, and
  Liu]{zhao2023chatbridge}
Zhao, Z., Guo, L., Yue, T., Chen, S., Shao, S., Zhu, X., Yuan, Z., and Liu, J.
\newblock Chatbridge: Bridging modalities with large language model as a
  language catalyst.
\newblock \emph{arXiv preprint arXiv:2305.16103}, 2023.

\end{thebibliography}

\newpage
\appendix
\onecolumn
\appendix

\section{Dataset Staging, Weights, and Templates}
\label{appendix: dataset}

Table \ref{tab: dataset by type} includes an overview of datasets (by type) we use to train Audio Flamingo. 
We construct instructions for each task and dataset. Below are all instruction templates we use.

\begin{tcolorbox}[colback=gray!5!white,colframe=gray!75!black]
  \textit{Audio Captioning:}
  
  \texttt{$\circ$ Describe the sound/music in a sentence.}

  \texttt{$\circ$ Describe the sound/music at length.}

  \texttt{ }
  
  \textit{Audio Question Answering:}

  \texttt{$\circ$ \{question\}}

  \texttt{$\circ$ Please answer this question: \{question\}}

  \texttt{$\circ$ Please answer this question: \{question\}. Options:\slashn- yes\slashn- no}

  \texttt{ }

  \textit{Audio Classification:}

  \texttt{$\circ$ Classify this sound. (Options: ...)}

  \texttt{$\circ$ Describe the sound in \{number\} words.}

  \texttt{$\circ$ What is the emotion of this speech? (Options: ...)}

  \texttt{$\circ$ What is the instrument/genre of this music? (Options: ...)}
  
  \texttt{$\circ$ This music note is produced by}
  
\end{tcolorbox}

The detailed pre-training datasets and their number of epochs are shown in Table \ref{tab: pretraining datasets}. The detailed SFT datasets and their number of epochs are shown in Table \ref{tab: sft datasets}. 

\begin{table*}[!h]
    \centering
    \caption{Pre-training datasets and epochs.}
    \begin{tabular}{cccc}
    \toprule
        Dataset & Audio Length & \#Audio-Text Pairs & Epochs \\ \hline
        OpenAQA & 693.2 hrs & 1959.8K & $1.0$ \\
        Laion630k\textsubscript{BBCSoundEffects} & 456.9 hrs & 15.1K & $5.0$ \\
        Laion630k\textsubscript{Freesound} & 2494.8 hrs & 306.5K & $1.0$ \\
        SoundDescs & 749.7 hrs & 23.1K & $1.0$ \\
        WavCaps & 3793.3 hrs & 402.6 K & $1.75$ \\
        AudioSet & 2617.8 hrs & 950.8K & $1.0$ \\
        WavText5K & 23.8 hrs & 4.3K & $3.0$ \\
        MSP-Podcast & 73.9 hrs & 45.1K & $1.2$ \\
        MELD & 8.7 hrs & 32.9K & $2.4$ \\
        MusicAVQA\textsubscript{audio-visual} & 142.4 hrs & 17.9K & $3.0$ \\
        MusicQA & 62.9 hrs & 70K & $1.2$ \\
        LP-MusicCaps\textsubscript{MSD} & 5805.7 hrs & 1331.8K & $1.0$ \\
        NSynth & 321.3 hrs & 289.2K & $0.4$ \\
        MTG-Jamendo & 3768.9 hrs & 55.6K & $1.0$ \\
    \bottomrule
    \end{tabular}
    
    \label{tab: pretraining datasets}
\end{table*}

\begin{table*}[!h]
    \centering
    \caption{SFT datasets and epochs.}
    \begin{tabular}{ccccc}
    \toprule
        Dataset & Audio Length & \#Audio-Text Pairs & Epochs & ICL Dataset Epochs \\ \hline
        ClothoAQA & 7.4 hrs & 9.7K & $3.5$ & $0.5$ \\
        OpenAQA & 693.2 hrs & 1959.8K & $0.1$ & - \\
        Clotho-v2  & 24.0 hrs & 19.2K & $2.0$ & $0.5$ \\
        Laion630k\textsubscript{Epidemic} & 209.4 hrs & 40.7K & $0.8$ & $0.2$ \\
        MACS & 10.9 hrs & 17.3K & $0.8$ & $0.2$ \\
        FSD50k & 80.8 hrs & 41.0K & $0.9$ & $0.3$ \\
        CochlScene & 169.0 hrs & 60.9K & $1.2$ & $0.3$ \\
        NonSpeech 7k & 6.2 hrs & 6.3K & $2.4$ & $0.6$ \\
        Chime-home & 5.0 hrs & 4.5K & $1.5$ & $0.5$ \\
        Sonyc-UST & 34.9 hrs & 27.9K & $0.8$ & $0.2$ \\
        Emov-DB & 7.8 hrs & 6.8K & $1.6$ & $0.4$ \\
        JL-Corpus & 1.4 hrs & 2.4K & $6.0$ & $1.5$ \\
        Tess & 1.6 hrs & 2.8K & $2.0$ & $0.5$ \\
        OMGEmotion & 3.0 hrs & 1.7K & $3.0$ & - \\
        MusicAVQA\textsubscript{audio-only} & 77.1 hrs & 5.7K & $5.0$ & $1.0$ \\
        MusicQA & 62.9 hrs & 70K & $0.35$ & $0.05$ \\
        LP-MusicCaps\textsubscript{MSD} & 5805.7 hrs & 1331.8K & $0.025$ & $0.007$ \\
        LP-MusicCaps\textsubscript{MTT} & 126.4 hrs & 46.9K & $0.8$ & $0.2$ \\
        LP-MusicCaps\textsubscript{MC} & 7.4 hrs & 7.9K & $2.0$ & - \\
        MusicCaps & 7.4 hrs & 2.6K & $6.0$ & - \\
        NSynth & 321.3 hrs & 289.2K & $1.0$ & $1.0$ \\
        MTG-Jamendo & 3768.9 hrs & 55.6K & $0.1$ & - \\
        MusDB-HQ & 29.1 hrs & 10.2K & $1.0$ & - \\
        FMA & 860.7 hrs & 104.2K & $0.4$ & $0.1$ \\
    \bottomrule
    \end{tabular}
    
    \label{tab: sft datasets}
\end{table*}

~~\newpage
\section{Generated dialogue datasets}
\label{appendix: dialogue}

\subsection{Overview}

In this section, we introduce methods to generate our \texttt{AF-Dialogue-AudioSetSL} and \texttt{AF-Dialogue-MusicCaps} datasets with GPT-4 \citep{achiam2023gpt}. 
\texttt{AF-Dialogue-AudioSetSL} is generated based on the annotated events and timestamps of strongly labeled AudioSet-SL \citep{gemmeke2017audio,hershey2021benefit}. There are 76k dialogues in the train split and 1.5k dialogues in test split. 
\texttt{AF-Dialogue-MusicCaps} is generated based on tags and descriptions of MusicCaps \citep{agostinelli2023musiclm}. There are 3.3k dialogues in the train split and 1.4k dialogues in the test split. 
Each training sample has one to four rounds (QA pairs), and each testing sample has three or four rounds.

\subsection{Prompts}
\label{sec: dialogue_prompts}
Below we show the prompts used to generate audio-based multi-turn dialogues from GPT-4 for the strongly labeled AudioSet-SL dataset~\cite{gemmeke2017audio,hershey2021benefit} and the MusicCaps dataset~\cite{agostinelli2023musiclm} respectively.


\begin{tcolorbox}[colback=gray!5!white,colframe=gray!75!black]

\textcolor{blue}{sys prompt:}~Based on the sound events, create a dialogue between you (the assistant) and a person (the user) about the events in the audio. Each dialogue should consist of: \\
1. A user examines the audio, and sends a reasonable and creative message to the assistant. 
2. Once the audio is provided, the assistant thoroughly perceives and comprehends them, responding with helpful answers that provide comprehensive reasoning. Do not include timestamps in the answer provided by the assistant.
3. Considering the past dialogue i.e. the question and the answer in the previous timestep, the user chooses another question or further inquiry.
4. The assistant provides a helpful answer providing comprehensive reasoning regarding the audio content.
\\
The audios are 10 seconds long and the sound events are segmented based on where they occur in the audio.
All the questions asked by the user should be diverse and complex.
\\

\textbf{Example 1:}
~Sound Events: Sound of Ambulance (siren) (High-pitched and wailing): [0.0s-1.0s]; Sound of Traffic noise, roadway noise (Droning, loud and intrusive): [0.0s-10.0s]; Sound of Accelerating, revving, vroom (Highpitched, short and intense): [2.0s-10.0s]; Sound of Generic impact sounds (Loud and sharp): [6.7s-6.8s]. 

\textbf{Dialogue:}
\\
User: What does the audio sound like?\\
Assistant: An ambulance siren echoes while traffic noise fades, and an engine revs.
\\
User: Can you describe in more detail about the engine revs?\\
Assistant: The engine revving is highpitched, short and intense. 
\\
User: From the highpitched sound, can you tell me what vehicle it was?\\ 
Assistant: It is hard to tell this from the information provided.
\\
User: Are there any other events in the audio that might be relevant?\\
Assistant: There is a sound of impact which is very loud and sharp.
\\

\textbf{Example 2:}
~
Sound Events: Sound of Outside, rural or natural: [0.000s-10.000s]; Sound of Liquid: [0.474s-0.680s], [4.249s-5.292s]; Sound of Gurgling, bubbling: [0.474s-0.680s], [4.249s-5.292s]; Sound of Splash, splatter: [1.620s-3.116s]; Sound of Laughter: [2.745s-4.029s], [5.299s-10.000s]; Sound of Chuckle, chortle: [2.752s-4.029s]; Sound of Speech: [4.551s-4.935s]; Sound of Male speech, man speaking: [4.557s-4.935s]; Sound of Belly laugh: [5.292s-10.000s]

\textbf{Dialogue:}
\\
User: What are the main events in the audio?\\
Assistant: Liquid sounds, laughter, speech, and a man speaking can be heard in a rural setting.
\\
User: Is is he speaking to other people?\\
Assistant: As there is sound of laughter present from other people, the man might be speaking to others.
\\
User: Can you explain in more detail about the first event?\\
Assistant: The liquid sounds are more of splashing, thrashing and splattering sounds.
\\
User: Are there other people laughing with him?\\
Assistant: Yes, there are other laughter voices along with that of the man.
\\

Based on the above examples, design a four-turn dialogue between the user and assistant for the example given below. 
The question asked by the user should use pronouns such as `he', `she', `it', `they' and `them' while referring to entities that have already occurred in the past questions or answers.
Format each QA pair in a single line as a JSON dictionary (key ``user'' for question, and ``assistant'' for answer, wrapped with \{ and \}). Do not include any other explanation.
\\
\textbf{Example 3:}\\

\end{tcolorbox}

\begin{tcolorbox}[colback=gray!5!white,colframe=gray!75!black]

\textcolor{blue}{sys prompt:}~
Based on the music, create a dialogue between you (the assistant) and a person (the user) about the events in the music audio. Each dialogue should consist of: \\
1. A user examines the music, and sends a reasonable and creative message to the assistant. 
2. Once the music audio is provided, the assistant thoroughly perceives and comprehends them, responding with helpful answers that provide comprehensive reasoning. Do not include timestamps in the answer provided by the assistant.
3. Considering the past dialogue i.e. the question and the answer in the previous timestep, the user chooses another question or further inquiry.
4. The assistant provides a helpful answer providing comprehensive reasoning regarding the music content.
The more complex and diverse the question, the better. You can use the list of questions below while asking questions. 
\\
\textbf{List of questions: 
}
Is it a live recording or a studio recording?\\
Is the guitar acoustic or electric?\\
Is it playing chords, a riff or a solo?\\
What music style or genre is it?\\
What key is it in? \\
How about the time signature?\\
What's the tempo? Slow, Medium or Fast?\\
Does it have vocals?    \\
Is it low pitched or high pitched?\\
Is the person speaking fast or slow?\\
What is the language and accent?\\
What emotion is conveyed by the voice?\\
What are the vocal characteristics of that emotion?\\
How old is the person likely to be given their vocal characteristics?\\

\textbf{Example 1:}
\\
Music description: House music with a thought provoking message in the beginning that eventually builds up and can be used to create the sense of a social revolution.
 
\textbf{Dialogue:}
\\
User: What type of music is this? \\
Assistant: It is house music with synth and drums and a man speaking at the beginning. 
\\
User: What is his tone while speaking?\\
Example 2: He has an intense and energetic tone in his voice. 
\\
User: Does he speak till the end? \\
Assistant: No, he stops speaking halfway and then it is followed by house music. 
\\
User: What is the source of this music?\\
Assistant: This is electronic music. 
\\
\textbf{Example 2: }
\\
Music description: Energetic bluesy song with a harmonica and horn section in musical dialogue.
\\
\textbf{Dialogue:}
\\
User: What instruments are playing in this music?\\
Assistant: This is a jazzy tune with harmonica, trumpet and drums playing. 
\\
User: Which key is the music playing in?\\
Assistant: The key is A Major. 
\\
User: How will you explain the tempo of this piece?\\
Assistant: The music is bouncy and positive with high beats per minute in the beginning and then changes signature halfway through.
\\
User: What is the source of this music?\\
Assistant: This is fusion jazz as it has elements of bass and synthesizer. 
\\
Based on above examples, design a four-turn dialogue between the user and assistant for the example given below. The question asked by the user should use pronouns such as `he', `she', 'it', 'they' and `them' while referring to entities that have already occurred in the past questions or answers.
Format each QA pair in a single line as a JSON dictionary (key ``user" for question, and ``assistant" for answer, wrapped with \{ and \}). Do not include any other explanation.
\\
\textbf{Example 3:}\\

\end{tcolorbox}

\newpage
\subsection{Dialogue filtering}
The dialogues generated by GPT-4 as discussed in \ref{sec: dialogue_prompts} do not always follow the prompts, resulting in answers that have phrases such as ``does not specify'', ``cannot be determined'', ``without additional context'' and so on. Hence, following~\citet{gardner2023llark}, we filter such QA pairs to improve the data quality and ensure desirable outputs from the model. 

\begin{figure}[!h]
    \centering
    \includegraphics[width=0.363\textwidth]{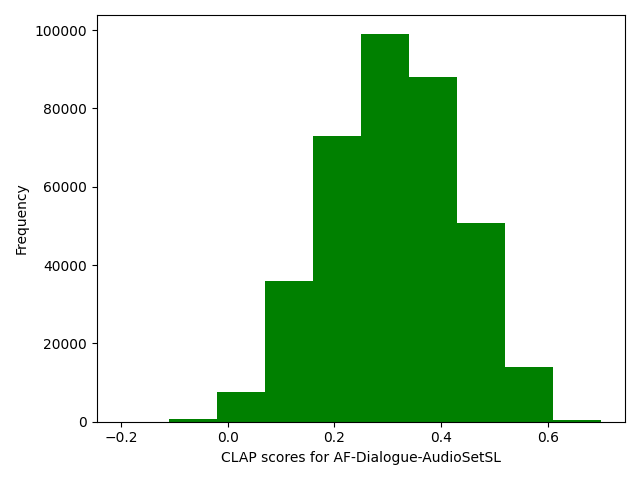}
    \includegraphics[width=0.4\textwidth]{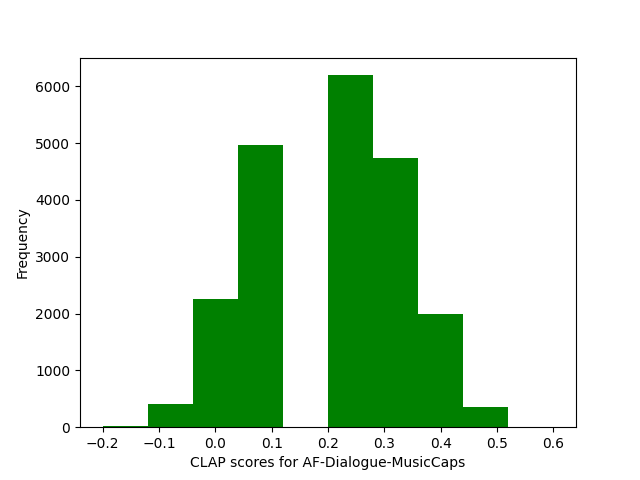}
    \caption{Distributions of LAION-CLAP similarities before filtering the generated \texttt{AF-Dialogue-AudioSetSL} (left) and \texttt{AF-Dialogue-MusicCaps} (right) datasets.}
    \label{fig: clap sim before filtering}
\end{figure}

Apart from the manual filtering step, we also filter samples based on the similarity of the answer generated by GPT-4 and the audio samples. Specifically, we compute the cosine similarity between the LAION-CLAP text-embeddings and audio-embeddings \citep{wu2023large} for a given QA pair in each dialogue. The distributions of similarities are shown in Figure~\ref{fig: clap sim before filtering}. We remove samples if the similarity is below a threshold of 0.3.

\newpage
\subsection{Examples of generated dialogues}

We demonstrate an example of generated dialogue for each of the generated dataset.

\begin{tcolorbox}[colback=gray!5!white,colframe=gray!75!black]



\texttt{[Audio annotations: Sound of Mechanisms (Produces sound through mechanical means.): [0.119s-10.000s]; Sound of Generic impact sounds (Short, sharp, and loud bursts.): [0.217s-0.387s], [2.394s-2.727s], [2.889s-3.255s], [3.799s-4.132s], [4.473s-4.766s], [4.928s-5.188s], [8.129s-10.000s]; Sound of Glass shatter (High-pitched, sharp, sudden, loud crash.): [1.622s-4.652s]; Sound of Walk, footsteps (Soft, rhythmic, echoing thuds.): [5.724s-6.106s], [6.455s-6.650s], [7.024s-7.479s], [7.625s-7.901s]]}

~~

User: Can you describe the main events in the audio?

Assistant: The audio mainly consists of mechanical sounds, sharp impact sounds, the shattering of glass, and the sound of footsteps.

User: What can you tell me about the impact sounds?

Assistant: The impact sounds are short, sharp, and loud bursts. They occur at several points throughout the audio, suggesting some kind of repetitive action or event.

User: What about the glass shattering, does it happen more than once?

Assistant: The glass shattering sound is a high-pitched, sharp, sudden, loud crash. It occurs once in the audio, lasting for about three seconds.

\end{tcolorbox}

\begin{tcolorbox}[colback=gray!5!white,colframe=gray!75!black]



\texttt{[Audio annotations: This is the recording of a cowbell. The cowbell is being played with the side and the tip of the drumstick repeatedly. This recording could be sampled for use in beat-making especially for genres that have a heavy usage of cowbells such as latin dance music.]}

~~

User: What instrument is being played in this audio?

Assistant: The instrument being played in this audio is a cowbell.

User: How is it being played?

Assistant: The cowbell is being played with the side and the tip of the drumstick repeatedly.

User: What is the tone of the cowbell?

Assistant: The tone of the cowbell is bright, resonant, and penetrating.

\end{tcolorbox}

~~\newpage
\section{Additional Experiments, Examples and Discussion of Audio Flamingo}
\label{appendix: demo}

We refer readers to visit our demo website: \url{https://audioflamingo.github.io/} for comprehensive demos.

\subsection{Additional Few-shot Experiments on Unseen Labels}
\label{appendix: additional few-shot}

Audio Flamingo can be adapted to classification with completely unseen labels with the help of ICL-based RAG. In Table \ref{tab: additional few-shot} we demonstrate Audio Flamingo can be adapted to new classification tasks where none of the labels have been seen during training. The labels are bird species and gun names, respectively. The huge improvement from zero-shot to few-shot indicates Audio Flamingo's strong ability to follow retrieved in-context samples. 

\begin{table}[!h]
    \centering
    \caption{Accuracy of Audio Flamingo on new classification tasks with unseen labels.}
    \begin{tabular}{ccc}
    \toprule
        Dataset & Zero-shot & Few-shot \\ \hline
        \texttt{BG-Gun-Sound} \citep{park2022enemy} & $1.6\%$ & $53.5\%$ \\
        \texttt{Birds-Dataset-Till-2022} \citep{birds2022} & $0.0\%$ & $42.2\%$ \\
    \bottomrule
    \end{tabular}
    \label{tab: additional few-shot}
\end{table}

\subsection{Compute}

In this section we compare the computation resources required by our model and baseline models. 
Pengi \citep{deshmukh2023pengi} is based on GPT2-base and has a total size 308M. It’s trained on 20 V100 GPUs.
LTU \citep{gong2023ltu} is based on LLaMA-7B and has a total size 6.74B. It’s trained on 4 RTX A6000 GPUs for about 3 days.
Qwen-Audio \citep{chu2023qwen} is based on Qwen-7B and has a total size of 8.3B. The training cost is unknown.
Salmonn \citep{tang2023salmonn} is based on Vicuna-13B. The training cost is unknown.
Our proposed Audio Flamingo has a total size of 2.2B. We use 8 A100 GPUs. It takes about 5-6 days for pretraining, about 3 days for SFT, and about 2hrs for dialogue fine-tuning.

\subsection{Audio Flamingo understands sound in depth}
\label{appendix: demo understanding}

\begin{tcolorbox}[colback=blue!5!white,colframe=gray!75!black]
\textbf{Audio Flamingo understands occurrences in audio.}

\texttt{[Subjective description of the audio: three people talking in order, with a lot of noise in the recording.]}

Task: \textit{Audio Question Answering}

Instruction: \textit{How many different voices are there?}

Ground Truth: \textit{Three.}

Baseline: \textit{One.}

Audio Flamingo: \textit{\underline{Three}.}
  
\end{tcolorbox}

\begin{tcolorbox}[colback=blue!5!white,colframe=gray!75!black]
\textbf{Audio Flamingo understands the order of sounds.}

\texttt{[Subjective description of the audio: a man speaks on a talk show, then the audience laugh and clap their hands.]}

Task: \textit{Audio Captioning}

Instruction: \textit{Describe this sound.}

Ground Truth: \textit{A man is speaking followed by applause.}

Baseline: \textit{A crowd of people are speaking and applauding.}

Audio Flamingo: \textit{A man is speaking \underline{and then} the audience claps.}
  
\end{tcolorbox}

\begin{tcolorbox}[colback=blue!5!white,colframe=gray!75!black]
\textbf{Audio Flamingo understands quality of audio.}

\texttt{[Subjective description of the audio: a beautiful and sad piece of instrumental music starting with keyboard playing chords slowly followed by soft violin.]}

Task: \textit{Audio Question Answering}

Instruction: \textit{What is the atmospheric quality of the audio?}

Ground Truth: \textit{The atmospheric quality of the audio is its ability to create a sense of space and depth, with a focus on creating a sense of mystery and wonder.}

Baseline: \textit{The audio has a rich and dynamic quality, which suggests a lively and engaging atmosphere, possibly associated with music or a musical performance, but it is not specified in the given audio events.}

Audio Flamingo: \textit{The audio has a \underline{melancholic and introspective atmosphere}.}
  
\end{tcolorbox}

\begin{tcolorbox}[colback=blue!5!white,colframe=gray!75!black]
\textbf{Audio Flamingo perceives loudness and its change over time.}

\texttt{[Subjective description of the audio: a man speaks in a small volume, then there is very loud burst, and then the man laughs for very short time in a surprised way.]}

Task: \textit{Audio Captioning}

Instruction: \textit{Describe the sound in a sentence.}

Ground Truth: \textit{A man briefly talks followed by a loud burst then laughter.}

Baseline: \textit{A person is laughing and a gun is fired.}

Audio Flamingo: \textit{A man is speaking and then a loud bang occurs.}
  
\end{tcolorbox}

\begin{tcolorbox}[colback=blue!5!white,colframe=gray!75!black]
\textbf{Audio Flamingo understands distances of sounds.}

\texttt{[Subjective description of the audio: a man speaks in the rain, then another man laughs very slightly, followed by the sound of thunderstorm.]}

Task: \textit{Audio Captioning}

Instruction: \textit{Describe this sound.}

Ground Truth: \textit{A man talking followed by another man laughing as rain falls and thunder roars in the distance.}

Baseline: \textit{A man is talking and rain is falling.}

Audio Flamingo: \textit{A man is talking in the rain with thunder \underline{in the background}.}
  
\end{tcolorbox}

\subsection{Audio Flamingo looks at retrieved samples}
\label{appendix: demo RAG}

\begin{tcolorbox}[colback=blue!5!white,colframe=gray!75!black]
\textbf{Audio Flamingo borrows key words from retrieved samples.}

Task: \textit{Audio Captioning}

Instruction: \textit{Describe this sound.}

Ground Truth: \textit{Paper is crinkling and crumpling, and thumping occurs.}

Baseline: \textit{Movement of a person in a kitchen with pots and pans on a counter.}

Retrieved samples:

\textit{1) Some material is \underline{crumpled}.}

\textit{2) Plastic is being \underline{crumpled}.}

\textit{3) \underline{Crinkling of a paper} with a small step.}

Audio Flamingo: \textit{A person \underline{crumples} up a sheet of \underline{paper} and \underline{crinkles} it.}

~~

Task: \textit{Audio Captioning}

Instruction: \textit{Describe this sound.}

Ground Truth: \textit{a woman speaks, and faucet water runs.}

Baseline: \textit{A woman is talking to a child.}

Retrieved samples:

\textit{1) \underline{A woman speaking} and \underline{water running}.}

\textit{2) A women voice followed by flowing water \underline{from a faucet}.}

\textit{3) Water sprays from a faucet, a person speaks.}

Audio Flamingo: \textit{\underline{A woman speaks} while \underline{water runs} \underline{from a faucet}.}
  
\end{tcolorbox}

\begin{tcolorbox}[colback=blue!5!white,colframe=gray!75!black]
\textbf{Audio Flamingo can ignore noisy retrieved samples and make correct prediction.}

Task: \textit{Music Genre Classification}

Instruction: \textit{What is the genre of this music?}

Ground Truth: \textit{pop.}

Retrieved samples:

\textit{1) reggae.}

\textit{2) jazz.}

\textit{3) disco.}

Audio Flamingo: \textit{pop.}

~~

Task: \textit{Music Instrument Classification}

Instruction: \textit{What is the instrument of this music?}

Ground Truth: \textit{flute.}

Retrieved samples:

\textit{1) clarinet.}

\textit{2) clarinet.}

\textit{3) clarinet.}

Audio Flamingo: \textit{flute.}

\end{tcolorbox}

\subsection{Audio Flamingo has strong dialogue ability}
\label{appendix: demo dialogue}

\begin{tcolorbox}[colback=blue!5!white,colframe=gray!75!black]

\texttt{[Audio annotations: Sound of Chirp, tweet (High-pitched, short, repetitive bird sounds.): [0.000s-1.157s], [1.543s-2.055s], [3.220s-4.409s], [4.748s-5.709s], [5.921s-6.780s], [7.094s-7.976s], [8.331s-8.819s], [9.142s-10.000s]; Sound of Animal (Loud, diverse, and often rhythmic.): [0.000s-1.157s], [1.528s-2.031s], [3.220s-4.386s], [4.764s-5.701s], [5.929s-6.740s], [7.071s-7.976s], [8.331s-8.795s], [9.134s-10.000s]; Sound of Bird vocalization, bird call, bird song (Melodic, rhythmic, varied pitch and tone.): [0.000s-1.165s], [1.528s-2.039s], [3.213s-4.386s], [4.772s-5.693s], [5.921s-6.701s], [7.047s-7.953s], [8.331s-8.780s], [9.150s-10.000s]; Sound of Bird (High-pitched, melodic, rhythmic chirping sounds.): [0.000s-1.173s], [1.535s-1.992s], [3.205s-4.362s], [4.780s-5.669s], [5.906s-6.669s], [7.016s-7.929s], [8.323s-8.772s], [9.134s-10.000s]; Sound of Speech (\"Sound waves produced by vocal cords.\"): [2.024s-3.008s], [4.630s-5.283s], [6.835s-8.134s], [8.701s-10.000s]; Sound of Female speech, woman speaking: [2.047s-3.039s], [4.614s-5.315s], [6.843s-8.150s], [8.709s-10.000s]]}

~~

User: \textit{Can you briefly explain what you hear in the audio?}

Ground Truth: \textit{The audio primarily consists of bird sounds, including chirping and bird vocalizations. There are also sounds of other animals. Additionally, there is a woman speaking at certain intervals.}

Baseline: \textit{Birds are chirping.}

Audio Flamingo: \textit{The audio features a variety of sounds including bird vocalizations, mechanical sounds, impact sounds, a man speaking, and a ticking sound.}

~~

User: \textit{Are the bird sounds continuous?}

Ground Truth: \textit{The bird sounds are not continuous. They occur at different intervals throughout the audio, with some overlap with the woman's speech.}

Baseline: \textit{Yes, the bird sounds are continuous.}

Audio Flamingo: \textit{No, the bird sounds are not continuous. They occur at different intervals throughout the audio.}

~~

User: \textit{Do the bird sounds and the person's speech overlap?}

Ground Truth: \textit{Yes, there are instances where the bird sounds and the woman's speech overlap.}

Baseline: \textit{Yes, the bird sounds and the woman's speech overlap.}

Audio Flamingo: \textit{Yes, the bird sounds and the person's speech do overlap at certain points in the audio.}

\end{tcolorbox}

\begin{tcolorbox}[colback=blue!5!white,colframe=gray!75!black]

\texttt{[Audio annotations: Sound of Music (Sound produced by vibrating instruments.): [0.000s-10.000s]; Sound of Rapping (Rhythmic speech over beats.): [0.000s-10.000s]; Sound of Male singing (Deep, resonant, and powerful tones.): [0.000s-10.000s]]}

~~

User: \textit{Is the male voice rapping or singing?
Ground Truth: The male voice is doing both, rapping and singing in the audio.}

Baseline: \textit{The male voice is rapping.}

Audio Flamingo: \textit{The male voice is rapping.}

~~

User: \textit{What kind of music is it?}

Ground Truth: \textit{The type of music isn't specified, but it's accompanied by rapping and male singing.}

Baseline: \textit{The kind of music is hip-hop.}

Audio Flamingo: \textit{The music is a rap song.}

~~

User: \textit{Is the male voice the only one present?}

Ground Truth: \textit{Yes, the male voice is the only one specified in the audio, both rapping and singing.}

Baseline: \textit{Yes, the male voice is the only one present.}

Audio Flamingo: \textit{Yes, the male voice is the only one present in the audio.}

\end{tcolorbox}

\begin{tcolorbox}[colback=blue!5!white,colframe=gray!75!black]

\texttt{[Audio annotations: Sound of Music (Sound produced by vibrating instruments.): [0.000s-10.000s]; Sound of Male singing (Deep, resonant, and powerful tones.): [1.341s-3.734s], [9.289s-10.000s],
]}

~~

User: \textit{What genre does this music belong to?}

Ground Truth: \textit{This music belongs to the R\&B/Disco genre.}

Baseline: \textit{It is not specified in the given audio clip.}

Audio Flamingo: \textit{This music belongs to the R\&B genre.}

~~

User: \textit{Can you tell me more about the vocals in this track?}

Ground Truth: \textit{Sure, the vocals in this track are from a male singer who is singing in a higher register. His voice has been modified with effects, making it sound like a voice sample.}

Baseline: \textit{The vocals in this track are described as rich and full-bodied, but no further details are given about the specific singer or the lyrics being sung in the audio clip.}

Audio Flamingo: \textit{The vocals in this track are by a male singer. His voice is deep, resonant, and powerful, adding to the overall emotional impact of the song.}
  
\end{tcolorbox}\label{sef:appendix}
\end{document}